\documentclass[12pt]{article}

\usepackage[a4paper,text={16.3cm,22cm}]{geometry}
\usepackage{amsmath,amsfonts,braket,slashed,amssymb,bm,bbm,xcolor,graphicx}
\usepackage[small,labelfont=bf]{caption}
\usepackage{subcaption}
\usepackage{cite}
\usepackage[
    bookmarksnumbered=true,
    urlcolor=blue,
    linkbordercolor=red,
    citebordercolor=green,
    bookmarksopen=true
    ]{hyperref}
\usepackage[section]{placeins}   
\usepackage{siunitx}
\usepackage{dsfont}

\allowdisplaybreaks
\numberwithin{equation}{section}

\newcommand{\argset}[1]{\{\underline{#1}\}}
\newcommand{\argsets}[2]{\{\underline{#1},\underline{#2}\}}
\newcommand{\als}{\alpha_s}
\newcommand{\tope}{\boldsymbol{T}}
\newcommand{\Coul}{{\text{Coul}}}
\newcommand{\gcusp}{\gamma_{\text{cusp}}}
\newcommand{\gcuspals}[1]{\gamma_{\text{cusp}}\bigl(\als({#1})\bigr)}

\DeclareMathOperator{\arcosh}{arcosh}
\DeclareMathOperator{\arcoth}{arcoth}
\DeclareMathOperator{\artanh}{artanh}
\DeclareMathOperator{\erf}{erf}
\DeclareMathOperator{\Tr}{Tr}

\title{Super-Leading Logarithms in Top-Quark\\[1ex] Pair Production at Hadron Colliders}

\begin{document}

\begin{titlepage}

\begin{flushright}
{\small
MITP-25-061\\
FERMILAB-PUB-25-0767-T-V\\
October 28, 2025}
\end{flushright}

\makeatletter
\vskip0.8cm
\pdfbookmark[0]{Super-Leading Logarithms in Top-Quark Pair Production at Hadron Colliders}{title}
\begin{center}
{\Large \bf\boldmath \@title}
\end{center}
\makeatother

\vspace{0.5cm}
\begin{center}
\textsc{Upalaparna Banerjee,$^{a,b}$ Romy Grünhofer,$^a$ Matthias König,$^a$ \\ Yibei Li,$^a$ Matthias Neubert,$^{a,c}$  
and Josua Scholze$^a$} \\[6mm]
    
\textsl{${}^a$PRISMA$^+$ Cluster of Excellence \& Mainz Institute for Theoretical Physics\\
Johannes Gutenberg University, Staudingerweg 9, 55128 Mainz, Germany\\[0.3cm]
${}^b$Particle Theory Department, Fermilab, Batavia, IL 60510, U.S.A.\\[0.3cm]
${}^c$Department of Physics \& LEPP, Cornell University, Ithaca, NY 14853, U.S.A.}
\end{center}

\vspace{0.6cm}
\pdfbookmark[1]{Abstract}{abstract}
\begin{abstract}
\!To date, the appearance and resummation of “super-leading” logarithms in hadron-hadron collisions has been studied only for massless parton states. We extend the formalism to include an arbitrary number of massive final states. We derive the corresponding anomalous dimension and identify an additional Coulomb phase that gives rise to a new source of super-leading logarithms.
We then perform a systematic leading-logarithmic resummation of these contributions for $2\to M$ processes.  Finally, we analyze the numerical impact in partonic scattering processes for $t\bar{t}$ production, including a treatment of the Sommerfeld enhancement observed near threshold.
\end{abstract}

\vfill\noindent\rule{0.4\columnwidth}{0.4pt}\\
\hspace*{2ex} {\small \textit{E-mail:} 
\href{mailto:ubanerjee@uni-mainz.de}{ubanerjee@uni-mainz.de}, 
\href{mailto:rgruenho@uni-mainz.de}{rgruenho@uni-mainz.de}, 
\href{mailto:matthias.koenig@uni-mainz.de}{matthias.koenig@uni-mainz.de},\\
\hspace*{2ex}\phantom{\textit{E-mail:} }
\href{mailto:yibli@uni-mainz.de}{yibei.li@uni-mainz.de}, 
\href{mailto:matthias.neubert@uni-mainz.de}{matthias.neubert@uni-mainz.de},
\href{mailto:j.scholze@uni-mainz.de}{j.scholze@uni-mainz.de}
}

\end{titlepage}

{\hypersetup{hidelinks}
\pdfbookmark[1]{Contents}{ToC}
\setcounter{tocdepth}{2}
\tableofcontents}
\vspace{6mm}

\section{Introduction}

Upcoming precision measurements at hadron colliders, such as the Large Hadron Collider (LHC) at CERN, demand increasingly accurate theory predictions for a wide range of collider observables. Among these, jet observables play a crucial role in improving our understanding of high-energy physics, as they closely mirror the underlying short-distance dynamics. Their theoretical description is challenging in part due to the presence of vetoes on soft radiation. In particular, such vetoes lead to the appearance of so-called non-global logarithms (NGLs)~\cite{Dasgupta:2001sh}, which arise from correlated soft emissions across measured and unmeasured regions of phase space. In hadron collisions, additional logarithmic corrections known as super-leading logarithms (SLLs)~\cite{Forshaw:2006fk,Forshaw:2008cq,Keates:2009dn} can occur, originating from a breakdown of color coherence and collinear factorization in space-like splittings due to Glauber-gluon exchanges between colored initial states~\cite{Catani:2011st,Forshaw:2012bi,Schwartz:2017nmr,Forshaw:2021fxs,Cieri:2024ytf,Guan:2024hlf,Henn:2024qjq,Becher:2024kmk}.

The resummation of the leading NGLs for gap-between-jets cross sections has been achieved in~\cite{Hatta:2013qj,Becher:2016mmh,Balsiger:2018ezi}, including the effects of massive final states~\cite{Balsiger:2020ogy}. These techniques were later extended to subleading logarithmic accuracy for massless partons~\cite{Banfi:2021owj,Banfi:2021xzn,Becher:2021urs,Becher:2023vrh}. At hadron colliders, on the other hand, SLLs represent the formally leading logarithmic corrections. They have been resummed at leading logarithmic accuracy for gap-between-jets partonic cross sections~\cite{Becher:2021zkk,Becher:2023mtx}, along with an arbitrary number of Glauber-gluon  insertions~\cite{Boer:2023jsy,Boer:2023ljq,Boer:2024xzy} and within renormalization-group (RG) improved perturbation theory~\cite{Boer:2024hzh}. A first phenomenological study of gap-between-jets hadronic cross sections has confirmed that SLLs can have sizable effects at low veto scales~\cite{Becher:2024nqc}.

Parton showers are another approach to resum large logarithmic corrections. However, traditional implementations rely on the large-$N_c$ approximation and therefore cannot capture SLLs, which enter only at subleading order in the $1/N_c$ expansion. To address this limitation, amplitude-level parton showers with exact color treatment and quantum interference are under active development, see e.g.~\cite{Nagy:2019pjp,Hoche:2020pxj,FerrarioRavasio:2023kyg,Forshaw:2025fif} and references therein.

In this work, we extend the discussion and resummation of SLLs to processes containing massive final states, focusing on the case of massless initial states. This is crucial for the precision physics program at the LHC, where heavy colored particles such as top quarks are produced abundantly. Our formalism will also be useful to study extensions of the Standard Model involving new heavy colored particles, such as squarks and gluinos or Kaluza-Klein gluons. 

This paper is structured as follows: in Section~\ref{sec:anomalous-dimension}, we derive the anomalous dimension of the hard functions with massive final-state partons. Section~\ref{sec:new-sources-of-SLLs} then demonstrates how this induces a new source of SLLs, and evaluates the relevant color structures and integrals, building on the work of~\cite{Becher:2023mtx,Boer:2024hzh}. In Section~\ref{sec:top_quark_pair_production}, we apply our results to $t\bar{t}$ production and provide the first numerical estimate of SLLs in partonic cross sections involving massive final states.

\section{Derivation of the anomalous dimension}
\label{sec:anomalous-dimension}

The resummation of SLLs is performed by means of a factorization theorem derived using effective field theory. For gap-between-jets observables at hadron colliders, the $2\to M$ production cross section can be written in a factorized form\footnote{The structure of the formula becomes more involved at three-loop order when contributions from subleading logarithms are included~\cite{Becher:2025igg}. These complications are beyond the scope of our work.} as~\cite{Balsiger:2018ezi,Becher:2021zkk,Becher:2023mtx}
\begin{align}\label{eq:factorization}
\sigma_{2\to M}(Q_0) =\int \!\! d\xi_1 \! \int \!\! d\xi_2 \!\!\sum_{m=2+M}^\infty \!\! \langle \boldsymbol{\mathcal{H}}_m (\argsets{n}{v},\argset{m},s,\xi_1,\xi_2,\mu) \otimes_{\!\int} \boldsymbol{\mathcal{W}}_m (\argsets{n}{v},Q_0,\xi_1,\xi_2,\mu) \rangle \,,
\end{align}
where $s$ is the squared center-of-mass energy of the hadronic system, $\xi_1,\xi_2$ denote the longitudinal momentum fractions of the two initial-state partons from the colliding hadrons, and $M$ denotes an arbitrary number of final-state partons involved in the hard scattering. The brackets $\langle\ldots\rangle$ denote a sum (average) over final-state (initial-state) color and spin indices. The soft parameter $Q_0$ sets the veto scale on soft radiation outside the jets. The set $\argsets{n}{v} = \{n_1,n_2, \dots,n_i,\dots,v_I,\dots\}$ collects the light-like directions $n_i=p_i/E_i$ (with $n_i^2=0$) and 4-velocities $v_I=p_I/m_I$ (with $v_I^2=1$) of the hard partons. Here and in the following, we use lowercase indices $(i, j,\dots)$ to denote massless particles, and uppercase indices $(I, J,\dots)$ for massive ones. Greek indices will be used in formulas that refer to both massive and massless partons.

The set $\argset{m} = \{m_1, \dots, m_m\}$ collects the masses of the $m$ hard partons. Note that depending on the context, $m$ denotes either the total parton multiplicity or a parton mass. Since our work focuses on the application to top-quark pair production, we assume that all masses are of order the hard scale set by the partonic center-of-mass energy $\sqrt{\hat s}$. The symbol $\otimes_{\!\int}$ indicates the integration over all final-state directions and over the energies of all massive partons
\begin{align}\label{eq:def_otimes_intMassive}
   \prod_{\alpha=3}^m \int [d\Omega_\alpha] \prod_{I}\int \frac{dE_I}{\tilde{c}^\varepsilon(2\pi)^2} (E_I^2-m_I^2)^{(d-3)/2} \theta(E_I-m_I)\,,
\end{align}
where the integral measure reads
 \begin{align}\label{eq:integral_measure_dOmegak}
     [d \Omega_\alpha] =\tilde{c}^\varepsilon \frac{d^{d-2}{\Omega_\alpha}}{2(2\pi)^{d-3}}\,,
 \end{align}
with $\tilde{c}=e^{\gamma_E}/\pi$ and $d=4-2\varepsilon$. Note that the integration is performed after multiplying $\boldsymbol{\mathcal{H}}_m$ and $\boldsymbol{\mathcal{W}}_m$, in contrast to the purely massless case. Pulling out the energy integrations from the hard functions is also necessary in some other scenarios~\cite{Becher:2023znt,Becher:2025igg}.

The hard function of multiplicity $m$ is defined as
\begin{align}\label{eq:hard_function}
    \boldsymbol{\mathcal{H}}_m=&
    \frac{1}{2 \xi_1 \xi_2 s} \prod_{i\neq1,2} \int \frac{dE_i \, E_i^{d-3}}{\tilde{c}^\varepsilon (2\pi)^2}   |\mathcal{M}_m(\argset{p}, \argset{m})\rangle \langle \mathcal{M}_m(\argset{p}, \argset{m})| \nonumber \\
    & \times (2\pi)^d \,2\,\delta(\bar{n}_1\cdot p_\text{tot}-\xi_1\sqrt{s}) \,\delta(\bar{n}_2\cdot p_\text{tot}-\xi_2\sqrt{s}) \,\delta^{(d-2)}(p^\perp_\text{tot}) \,\Theta_\text{hard} (\argsets{n}{v})\,,
\end{align}
where the energy integration is only performed for massless partons. The total momentum is given by $p_\text{tot}$, and $p^\perp_\text{tot}$ denotes the total momentum perpendicular to the beam directions $n_1$ and $n_2$. The angular constraint $\Theta_\text{hard}$ excludes hard partons from the veto region. The low-energy matrix elements $\boldsymbol{\mathcal{W}}_m$ involve soft emissions from the hard partons, accounted for by soft Wilson lines $S_{n_i}$ ($S_{v_I}$) for massless (massive) partons, as well as the non-perturbative collinear physics associated with the initial-state particles.

In order to achieve the resummation, $\boldsymbol{\mathcal{H}}_m$ and $\boldsymbol{\mathcal{W}}_m$ in \eqref{eq:factorization} are RG-evolved to a common scale. We choose the soft scale $Q_0$ because, at leading order, the low-energy matrix elements then reduce to products of parton distribution functions. To evolve the hard functions to the soft scale, we derive the anomalous dimension $\boldsymbol{\Gamma}^{\mathcal{H}}$ of the hard functions in the presence of massive final states. At the amplitude level, the anomalous dimension involving massive and massless partons is already known~\cite{Becher:2009qa,Becher:2009kw}, but has to be reorganized for our purpose. Following the strategy of~\cite{Becher:2023mtx}, we analyze the soft and collinear limits to extract all infrared (IR) divergences. We consider the contributions from virtual- and real-emission corrections separately.

\subsection{Divergences in virtual corrections}
\label{subsec:divergences-virtual-corrections}

\subsubsection*{Anomalous dimension of the amplitude}
The IR singularities of QCD amplitudes involving arbitrary numbers of massive and massless partons can be absorbed into a multiplicative renormalization factor. The corresponding anomalous-dimension matrix is given by~\cite{Becher:2009kw,Ferroglia:2009ep,Ferroglia:2009ii}
\begin{align}\label{eq:anomalous_dimension_M}
    \boldsymbol{\Gamma}^{\mathcal{M}}(\argset{p}, \argset{m}, \mu) &= \frac{1}{2} \sum_{(ij)} \boldsymbol{T}_i\cdot\boldsymbol{T}_j \gcusp(\als) \ln\!\left(\frac{\mu^2}{-s_{ij}}\right) + \sum_i \gamma^i(\als) \nonumber \\
    &\quad - \frac{1}{2} \sum_{(IJ)} \boldsymbol{T}_I\cdot\boldsymbol{T}_J \gcusp (\beta_{IJ},\als)  + \sum_I \gamma^I(\als) \nonumber \\
    &\quad+ \sum_{I,j} \boldsymbol{T}_I \cdot \boldsymbol{T}_j \gcusp(\als) \ln\!\left(\frac{ m_I \mu}{-s_{Ij}}\right) + \mathcal{O}(\als^2)\,.
\end{align}
The notation $(\alpha\beta)$ in the summation refers to all pairs of distinct parton indices $\alpha\neq \beta$, where both orderings $\alpha,\beta$ and $\beta,\alpha$ are included. The Lorentz-invariant quantity $s_{\alpha\beta}$ is defined as $s_{\alpha\beta} = 2 \sigma_{\alpha\beta} \, p_\alpha \cdot p_\beta + i 0$, where $\sigma_{\alpha\beta}$ equals $+1$ if partons $\alpha,\beta$ are both incoming or both outgoing, and $-1$ otherwise. 
$\boldsymbol{T}_\alpha$ specifies a color generator acting on particle $\alpha$ in the color-space formalism~\cite{Catani:1996vz}.

In the terms involving two massive partons in \eqref{eq:anomalous_dimension_M}, the cusp anomalous dimension $\gcusp$ depends explicitly on the cusp angle $\beta_{IJ}$, formed by the time-like Wilson lines associated with the partons $I$ and $J$. The cusp angle is determined by the parton masses and the momentum transfer through
\begin{align}\label{eq:betaIJ}
    \beta_{IJ} = \arcosh\!\left(\frac{-s_{IJ}}{2 m_I m_J}\right).
\end{align}
For the purpose of resumming the leading logarithms, we need the following expansions of the anomalous-dimension coefficients:
\begin{align}
   \gcusp(\als) 
   &= \frac{\als}{4\pi}\,\gamma_0 + \left( \frac{\als}{4\pi} \right)^2 \gamma_1
    + \mathcal{O}(\als^3) \,, \nonumber\\
   \gcusp(\beta_{IJ}, \als) 
   &= \frac{\als}{4\pi}\,\gamma_0\,\beta_{IJ} \coth(\beta_{IJ}) + \mathcal{O}(\als^2) \,, 
    \nonumber\\
   \gamma^{\alpha}(\als)  
   &= \frac{\als}{4\pi}\,\gamma_0^\alpha + \mathcal{O}(\als^2) \,,
\end{align}
where \cite{Becher:2009cu,Becher:2009kw} 
\begin{align}\label{eq:gammacoefs}
   \gamma_0 &= 4 \,, \hspace{1.8cm}
   \gamma_1 = \left( \frac{268}{9} - \frac{4\pi^2}{3} \right) C_A 
    - \frac{80}{9} T_F n_f \,, \nonumber\\
   \gamma^q_0 &= -3C_F \,, \qquad
   \gamma^Q_0 = -2C_F \,, \qquad
   \gamma^g_0 = -\frac{11}{3}C_A +\frac{4}{3}T_F n_f = - \beta_0 \,.
\end{align}
Here $C_F=(N_c^2-1)/(2N_c)$, $C_A=N_c$, $T_F=1/2$, $n_f$ denotes the number of light quark flavors, and $\beta_0$ is the first coefficient of the QCD $\beta$-function.

To compute the hard functions $\boldsymbol{\mathcal{H}}_m$, one needs to consider the squared amplitude, the relevant energy integrations and phase-space constraints \eqref{eq:hard_function}. From the structure of $\boldsymbol{\Gamma}^\mathcal{M}$ in \eqref{eq:anomalous_dimension_M}, it is evident that the cusp logarithms depend on the parton energies through the invariants $s_{\alpha\beta}$. In order to connect $\boldsymbol{\Gamma}^{\mathcal{M}}$ to the anomalous dimension of $\boldsymbol{\mathcal{H}}_m$, it is necessary to separate the energy dependence from the angular dependence in these logarithms, i.e.\ 
\begin{align}
    s_{ij} = 2\sigma_{ij} E_i E_j\,n_i\cdot n_j + i0 \,, \qquad
    s_{Ij} = 2\sigma_{ij} m_I E_j\,v_I\cdot n_j + i0 \,.
\end{align}
Doing so makes the imaginary parts of the cusp logarithms explicit, which are non-zero only when the partons involved are both incoming or outgoing. These phases originate from Glauber-gluon exchange and are commonly referred to as Glauber phases. 

The splitting of the logarithms also makes it possible to extract the purely soft contribution from the soft-collinear part of the anomalous dimension $\boldsymbol{\Gamma}^{\mathcal{M}}$. Specifically, the angular dependence of the cusp logarithms can be expressed as an integral over the directions of virtual gluons, where the integrand is given by the soft dipole operator with their collinear limits subtracted.
For two massive partons, the relevant part of $\boldsymbol{\Gamma}^{\mathcal{M}}$ can be decomposed into real and imaginary parts as
\begin{align}\label{eq:split_logarithms_both_massive}
    \beta_{IJ} \coth \beta_{IJ}
    = \left( \tilde{\beta}_{IJ} - i \pi \right) \frac{v_I \cdot v_J}{\sqrt{(v_I \cdot v_J)^2-1}}
\end{align}
with $\tilde{\beta}_{IJ} = \ln(v_I \cdot v_J+\sqrt{(v_I \cdot v_J)^2-1})$.
Analogous to the massless case discussed in \cite{Becher:2023mtx}, the logarithms in \eqref{eq:anomalous_dimension_M} can be identified as angular integrals over dipole radiator functions sourced by the corresponding partons~\cite{Lyubovitskij:2021ges}, i.e.
\begin{align}\label{eq:rewriting_GammaM_logarithms_kinematical_dependence}
     \int [d\Omega_k] \,\overline{W}_{ij}^k 
     &= - \ln\!\left(\frac{2}{n_i \cdot n_j}\right) + \mathcal{O}(\varepsilon) \,, 
      \nonumber\\
    \int [d\Omega_k] \,\overline{W}_{Ij}^k 
    &= - \ln\!\left(\frac{1}{v_I \cdot n_j}\right) + \mathcal{O}(\varepsilon) \,, 
     \nonumber\\
    \int [d\Omega_k] \,\overline{W}_{IJ}^k  
    &= \tilde{\beta}_{IJ} \frac{v_I \cdot v_J}{\sqrt{(v_I \cdot v_J)^2-1}} 
     + \mathcal{O}(\varepsilon) \,.
\end{align}
The soft dipole radiators are expressed in terms of the 4-velocities and light-like directions of the partons, and the reference vectors $n_k$ are aligned with the directions of the virtual gluons, so that
\begin{align}\label{eq:subtracted_dipoles}
     \overline{W}_{ij}^k &= \frac{n_i \cdot n_j}{n_i \cdot n_k n_j \cdot n_k} - \frac{1}{n_i \cdot n_k}\delta(n_i - n_k) - \frac{1}{n_j \cdot n_k}\delta(n_i - n_k)\,, \nonumber \\
    \overline{W}_{Ij}^k &= \frac{v_I \cdot n_j}{v_I \cdot n_k n_j \cdot n_k} - \frac{1}{n_j \cdot n_k}\delta(n_j - n_k)\,, \nonumber \\
    \overline{W}_{IJ}^k &= \frac{v_I \cdot v_J}{v_I \cdot n_k \, v_J \cdot n_k}\,.
\end{align}
Note that due to the light-like reference vectors, the radiator functions for massless partons need to have distinct indices. This also defines the subtracted massless monopoles $\overline{W}_{ii}^k$ to be zero. For the massive partons, however, the monopole functions $\overline W_{II}^k$ are non-zero.

Each contribution with explicit energy dependence in \eqref{eq:anomalous_dimension_M} involves only the energy of a single massless parton and thus can be simplified using color conservation,
\begin{align}\label{eq:rewriting_GammaM_logarithms_energy_dependence}
    \frac{1}{2} \sum_{(ij)} \boldsymbol{T}_i \cdot \boldsymbol{T}_j \left[\ln\!\left(\frac{\mu}{2 E_i}\right) + \ln\!\left(\frac{\mu}{2 E_j}\right)\right] 
    + \sum_{I,j} \boldsymbol{T}_I \cdot \boldsymbol{T}_j \ln\!\left(\frac{\mu}{2 E_j}\right) = - \sum_j C_j \ln\!\left(\frac{\mu}{2 E_j}\right) ,
\end{align}
where $C_j$ denotes the quadratic Casimir corresponding to the representation of the parton $j$.
The imaginary part of $\boldsymbol{\Gamma}^{\mathcal{M}}$ can be simplified using color conservation as well. We use the notation $2\Pi_{\alpha\beta}=\sigma_{\alpha\beta}+1$ and find that the additional color structures relative to the massless case involve only massive partons,
\begin{align}\label{eq:rewriting_GammaM_imaginary_part}
    &\frac{1}{2}\sum_{(ij)} \boldsymbol{T}_i\cdot\boldsymbol{T}_j \Pi_{ij} + \sum_{I,j} \boldsymbol{T}_I \cdot \boldsymbol{T}_j \Pi_{Ij} - \frac{1}{2} \sum_{(IJ)} \boldsymbol{T}_I \cdot \boldsymbol{T}_J \left(-\frac{v_I\cdot v_J}{\sqrt{(v_I \cdot v_J)^2-1}}\right) \nonumber \\
    &\qquad = 2 \boldsymbol{T}_1 \cdot \boldsymbol{T}_2  + \frac{1}{2} \left( C_1 + C_2 - \sum_{\alpha\neq1,2} C_\alpha\right) + \frac{1}{2}\sum_{(IJ)} \boldsymbol{T}_I \cdot \boldsymbol{T}_J \left(\frac{(v_I \cdot v_J)}{\sqrt{(v_I \cdot v_J)^2-1}}- 1 \right).
\end{align}
From now on, we will denote the kinematical factor in the Coulomb phase by 
\begin{align}
    v_{IJ} \equiv \frac{(v_I \cdot v_J)}{\sqrt{(v_I \cdot v_J)^2-1}}- 1
\end{align}
for notational simplicity.
Combining the contributions from \eqref{eq:rewriting_GammaM_logarithms_kinematical_dependence}, \eqref{eq:rewriting_GammaM_logarithms_energy_dependence}, and \eqref{eq:rewriting_GammaM_imaginary_part}, the full anomalous dimension of the amplitude $|\mathcal{M}_m\rangle$ can be written with the angular and energy dependence explicitly separated out,
\begin{align}\label{eq:rewriting_GammaM_result}
    \boldsymbol{\Gamma}^{\mathcal{M}}(\argset{p}, \argset{m}, \mu) = \frac{\als}{4\pi} \Bigg\{&- \frac{1}{2} \gamma_0 \sum_{(\alpha \beta)} \boldsymbol{T}_\alpha \cdot \boldsymbol{T}_\beta \int \frac{d^2\Omega_k}{4\pi} \overline{W}_{\alpha\beta}^k - \gamma_0\sum_i C_i \ln\!\left(\frac{\mu}{2 E_i}\right) \nonumber \\
    & + i \pi \gamma_0 \Bigg[2 \boldsymbol{T}_1\cdot\boldsymbol{T}_2 + \frac{1}{2}\sum_{(IJ)} \boldsymbol{T}_I \cdot \boldsymbol{T}_J v_{IJ} \nonumber \\
    &\qquad+ \frac{1}{2}\left(C_1 + C_2 - \sum_{\alpha\neq1,2} C_\alpha\right) \Bigg] + \sum_{\alpha} \gamma_0^\alpha\Bigg\} +\mathcal{O}(\als^2)\,.
\end{align}
Here, we introduced the compact notation
\begin{align}\label{eq:sum_alphabeta_definition}
    \sum_{(\alpha \beta)}\equiv\sum_{(i j)} + \sum_{(I J)} + 2 \sum_{I,j}
\end{align}
to unify the sum over the integrated dipoles, taking into account the double counting that is present in the terms involving two massive or two massless partons.

\subsubsection{Divergences in the hard functions}

We are now in a position to relate the divergences of the amplitude to those of the hard functions $\boldsymbol{\mathcal{H}}_m$. Since $\boldsymbol{\mathcal{H}}_m$ is related to the squared amplitude via \eqref{eq:hard_function}, the renormalization constant obtained from $\boldsymbol{\Gamma}^{\mathcal{M}}$ in \eqref{eq:rewriting_GammaM_result} determines the renormalization constant of the hard functions
\begin{align}\label{eq:ZH_expressed_via_ZM}
    (\boldsymbol{Z}^\mathcal{H})^{-1}(\argset{p}, \argset{m}, \mu, \varepsilon) = (\boldsymbol{Z}^\mathcal{M})^{-1}_L(\argset{p}, \argset{m}, \mu, \varepsilon) \, ((\boldsymbol{Z}^\mathcal{M})^{-1}_R)^\dag(\argset{p}, \argset{m}, \mu, \varepsilon)\,,
\end{align}
where the index $R$ ($L$) denotes that the corresponding renormalization constant has been derived from the divergences of the (conjugate) amplitude. The operators act from the right (left) on the hard functions.

In order to combine virtual and real contributions in Section~\ref{subsec:anomalous-dimension-hard-function}, we include the massive monopoles $\overline{W}_{II}^k$ in the first line of \eqref{eq:rewriting_GammaM_result}, which exactly corresponds to the terms involving the collinear anomalous dimension $\sum_I \gamma_0^I$ for massive partons in the last line. Hence, we obtain the following divergences from virtual contributions in the hard functions:
\begin{align}\label{eq:virtual_divergences_hard_function}
    {\boldsymbol{\mathcal{H}}}_m(\argsets{n}{v}, \!\argset{m}, \varepsilon) \!\supset &\frac{\als}{4\pi}\Bigg\{\! -\!\frac{\gamma_0}{4\varepsilon} \sum_{\alpha, \beta} \left(\boldsymbol{T}_{\alpha,L} \cdot \boldsymbol{T}_{\beta,L} \hspace{-1.1pt}+\hspace{-1.1pt} \boldsymbol{T}_{\alpha,R} \cdot \boldsymbol{T}_{\beta,R} \right) \! \int \!\frac{d^2\Omega_k}{4\pi} \overline{W}_{\alpha\beta}^k {\boldsymbol{\mathcal{H}}}_m(\argsets{n}{v}, \!\argset{m}, \mu) \nonumber \\
    &\quad + \frac{\gamma_0}{\varepsilon} i \pi \bigg[ \left(\boldsymbol{T}_{1,L} \cdot \boldsymbol{T}_{2,L} - \boldsymbol{T}_{1,R} \cdot \boldsymbol{T}_{2,R} \right) \nonumber\\
    &\qquad \qquad \quad + \frac{1}{4}\sum_{(IJ)} \left(\boldsymbol{T}_{I,L} \cdot \boldsymbol{T}_{J,L} - \boldsymbol{T}_{I,R} \cdot \boldsymbol{T}_{J,R}\right) v_{IJ} \bigg] {\boldsymbol{\mathcal{H}}}_m(\argsets{n}{v}, \!\argset{m}, \mu) \nonumber \\
    &\quad -\int \! d\mathcal{E}_m  \sum_i \left[\gamma_0 C_i \left(\frac{1}{2\varepsilon^2}+\frac{1}{\varepsilon} \ln\!\left(\frac{\mu}{2E_i}\right)\right) -  \frac{\gamma_0^i}{\varepsilon}\right] \widetilde{\boldsymbol{\mathcal{H}}}_m(\argset{p}, \!\argset{m}, \mu) \Bigg\}\,,
\end{align}
where $\int\! d\mathcal{E}_m$ represents the massless final-state energy integrations with the phase-space constraints and flux factor from~\eqref{eq:hard_function}, and $\widetilde{\boldsymbol{\mathcal{H}}}_m$ are the associated integrands of the hard functions~\cite{Becher:2023mtx}, i.e., the squared hard-scattering amplitudes. The first line of this expression now includes the massive monopoles. As in the massless case, the collinear terms in the last line involve only the massless partons. We can see that the constant imaginary part of $\boldsymbol{\Gamma}^{\mathcal{M}}$ drops out of the anomalous dimension of $\boldsymbol{\mathcal{H}}_m$, while an additional Coulomb phase (third line) appears due to the presence of massive final-state particles.

\subsection{Divergences in real emissions}
\label{subsec:divergences-real-emissions}

Similar to the virtual corrections discussed in the previous section, real emissions also lead to divergences when the emitted gluon becomes either soft or collinear to another hard parton involved in the process. However, unlike virtual corrections, real-emission divergences appear from integrating over the phase space of unresolved real particles. In this section, we focus on the structure of these divergences in the real-emission contributions and how they factorize in soft and collinear limits.

\subsubsection{Soft limit}

Analogous to the massless case, the amplitude $ |\mathcal{M}_{m+1}\rangle$ with $m$ hard partons of momenta $\{\underline{p}\}$ and one additional gluon of momentum $q$ factorizes in the limit where the gluon becomes soft~\cite{Catani:1996vz},
\begin{align}\label{eq:factorization_M_soft_emission}
    |\mathcal{M}_{m+1}^{a_q}(\{\underline{p},q\}, \argset{m})\rangle = \varepsilon^{*}_\mu \boldsymbol{J}^{\mu, a_q} |\mathcal{M}_m(\argset{p}, \argset{m}\rangle = \varepsilon_\mu^*(q; s_q) g_s \sum_\alpha \frac{-p_\alpha^\mu}{p_\alpha \cdot q} \boldsymbol{T}_\alpha^{a_q} |\mathcal{M}_m(\argset{p}, \argset{m}\rangle\,.
\end{align}
Here, $\varepsilon^{*}_\mu$ is the polarization vector of the emitted gluon, $s_q$ and $a_q$ denote its spin and color index, and $\boldsymbol{J}^{\mu,a_q}$ is the soft current, which we express at leading order in $g_s$.

Starting with the expression for the hard function involving $(m+1)$ partons, we apply soft factorization to take the soft limit. Carrying out the polarization sum over the emitted gluon, we then get the leading contribution
\begin{align}\label{eq:hard_function_soft_emission_before_dipole_substitution}
    &\sum_{s_q} \boldsymbol{\mathcal{H}}_{m+1}(\{\underline{n},\underline{v},n_q\}, \argset{m}) \nonumber \\
    &\qquad= -\int \frac{d E_{q}\,E_{q}^{d-3}}{\tilde{c}^\varepsilon(2\pi)^2} g_s^2  \sum_{\alpha,\beta} \frac{p_\alpha \cdot p_\beta}{(p_\alpha \cdot q)(p_\beta \cdot q)}  \theta_{\text{hard}}(n_q) \boldsymbol{T}_\alpha^{a_q} \boldsymbol{\mathcal{H}}_m(\argsets{n}{v}, \argset{m}) \boldsymbol{T}_\beta^{\tilde{a}_q}\,,
\end{align}
where $a_q$ and $\tilde{a}_q$ are the color indices of the additional gluon. For the phase-space constraints, the soft momentum $ q $ can be neglected in the $\delta$-functions, and the angular constraint factorizes as 
$\Theta_{\text{hard}}(\{\underline{n},\underline{v}, n_q\}) = \Theta_{\text{hard}}(\argsets{n}{v})\, \theta_{\text{hard}}(n_q)$. This allows us to express $ \boldsymbol{\mathcal{H}}_{m+1} $ in terms of $ \boldsymbol{\mathcal{H}}_m $. The overall negative sign arises because only the $ (-g^{\mu\nu}) $ term in the polarization sum contributes. The remaining terms are proportional to $ \sum_\alpha \boldsymbol{T}_\alpha $ and vanish due to color conservation.

Substituting the 4-momenta $ q = E_q\,n_q $, $ p_i = E_i\,n_i $, and $ p_I = m_I\,v_I $ as appropriate, we identify the dipole structures $ W_{\alpha \beta}^q $ in \eqref{eq:hard_function_soft_emission_before_dipole_substitution}. 
Since our focus here is on purely soft singularities, we replace $ W_{\alpha \beta}^q $ with its soft part $ \overline{W}_{\alpha \beta}^q $. To isolate the soft divergence, we introduce a cutoff on $E_q$ following~\cite{Becher:2021urs}.  
Evaluating the integral over $ E_q $, we extract the divergence
\begin{align}\label{eq:hard_function_soft_emission_divergence}
    \sum_{s_q} \boldsymbol{\mathcal{H}}_{m+1}(\{\underline{n},\underline{v},q\}, \argset{m}, \varepsilon) \supset \frac{\als}{\pi} \frac{1}{2\varepsilon}  \sum_{\alpha,\beta} \theta_{\text{hard}}(n_q) \overline{W}_{\alpha \beta}^q   \boldsymbol{T}_{\alpha,L} \circ \boldsymbol{T}_{\beta,R} \boldsymbol{\mathcal{H}}_m(\argsets{n}{v}, \argset{m}, \mu)\,.
\end{align}
We follow the notion of~\cite{Becher:2023mtx} to define the symbol $\circ$ as $\boldsymbol{T}_{\alpha,L} \circ \boldsymbol{T}_{\beta,R} \boldsymbol{\mathcal{H}}_m = \boldsymbol{T}_\alpha^{a_q} \boldsymbol{\mathcal{H}}_m\boldsymbol{T}_\beta^{\tilde{a}_q}$, which implies that the color basis is extended to include the emitted gluon. Note that we again obtain a massive monopole contribution, whereas the remaining structure is identical to the massless case.

\subsubsection{Collinear limit}
\label{subsubsec:collinear-limit}

We now consider the limit in which two partons, $ \alpha $ and $ \beta $, become collinear~\cite{Catani:1999ss,Catani:2003vu,Catani:2011st,Becher:2009qa}. This can be described as the collinear splitting of a parent parton $ P $, either time-like, $ P \to \alpha + \beta $, with both $ \alpha $ and $ \beta $ being final-state partons, or space-like, $ \alpha \to P + \beta $, corresponding to the splitting of an initial-state parton. In this subsection, we consider possible additional sources of collinear divergences that are produced in the presence of massive partons. The splittings involving only massless partons were discussed in~\cite{Becher:2023mtx}.

Since the initial-state partons are considered here to be massless, the space-like splitting does not introduce any new divergences compared to the massless case. In particular, the splitting of a gluon into two massive quarks cannot be a space-like collinear splitting because the far off-shell gluon would render the intermediate-state quark a part of the hard scattering.

As in the soft limit, the amplitude factorizes in the time-like collinear limit. The product of the hard function and the low-energy matrix element then becomes 
\begin{align}\label{eq:collinear_splitting_timelike_general_equation}
    &\langle \boldsymbol{\mathcal{H}}_{m+1}(\argsets{n}{v},\argset{m},\varepsilon) \otimes_{\!\int} \boldsymbol{\mathcal{W}}_{m+1}(\argsets{n}{v},\varepsilon) \rangle \nonumber \\
    &\qquad = \langle \prod_{\gamma=3}^{m+1} \int [d\Omega_\gamma] \int\hspace{-9pt}\int\! d \mathcal{E}_{m+1}  4\pi \als \tilde{\mu}^{2\varepsilon} \frac{2}{m_\alpha^2 + m_\beta^2 + 2 p_\alpha \cdot p_\beta - m_P^2} \mathcal{P}_{\alpha+\beta \leftarrow P}(\xi) \nonumber \\
    &\qquad \qquad \times\widetilde{\boldsymbol{\mathcal{H}}}_{m}(\argsets{\hat{n}}{\hat{v}},\argset{\hat{m}},\mu)  \boldsymbol{\mathcal{W}}_{m}(\argsets{\hat{n}}{\hat{v}},\mu)  \rangle\,,
\end{align}
where the sets $\argsets{\hat{n}}{\hat{v}}$ and $\argset{\hat{m}}$ include the parent parton $P$ instead of the collinear partons $\alpha$ and $\beta$. The scale $\tilde\mu$ is defined by $\tilde{\mu}^2=\mu^2 e^{\gamma_E}/(4\pi)$. The double integral sign implies an integration over all final-state energies, now including the massive final states, see~\eqref{eq:def_otimes_intMassive}. The functions $\mathcal{P}_{\alpha+\beta \leftarrow P}(\xi)$ are so-called splitting functions and depend on the momentum split defined by $p_\alpha = \xi P$, $p_\beta = (1-\xi) P$. Additionally, they depend on the masses as well as on the product $p_\alpha\cdot p_\beta$, which we omit in our notation. For the different possible splittings, the massive splitting functions read~\cite{Catani:2000ef}
\begin{align}\label{eq:splitting_functions}
    \mathcal{P}_{q + g \leftarrow q}(\xi) &= \mathcal{P}_{\Bar{q} + g \leftarrow \Bar{q}}(\xi) = C_F \left[\frac{1+\xi^2}{1-\xi} - \varepsilon(1-\xi) - \frac{m_q^2}{p_q \cdot p_g}\right], \nonumber \\
    \mathcal{P}_{g + q \leftarrow q}(\xi) &= \mathcal{P}_{ g + \Bar{q}  \leftarrow \Bar{q}}(\xi) = \mathcal{P}_{q + g \leftarrow q}(1-\xi)\,, \nonumber\\
    \mathcal{P}_{q + \Bar{q} \leftarrow g}(\xi) &= \mathcal{P}_{\Bar{q} + q  \leftarrow g}(\xi) = T_F \left[1- \frac{2\xi(1-\xi)}{1-\varepsilon} + \frac{1}{1-\varepsilon} \frac{m_q^2}{m_q^2 + p_q \cdot p_{\Bar{q}}}\right].
\end{align}
For massive partons, this is referred to as the quasi-collinear limit.

We now rewrite the energy integration in \eqref{eq:collinear_splitting_timelike_general_equation} using the momentum fraction~$\xi$,
\begin{align}\label{eq:energy_integrals_rewritten_collinear_splitting}
    \int\hspace{-9pt}\int\! d \mathcal{E}_{m+1} &= \int\hspace{-9pt}\int\! d \mathcal{E}_m \int_0^1 d\xi \frac{E_P}{\tilde{c}^\varepsilon (2\pi)^2} 
    \left\{\frac{(\xi^2E_P^2-m_\alpha^2)[(1-\xi)^2E_P^2-m_\beta^2]}{E_P^2-m_P^2}\right\}^{(d-3)/2} \nonumber \\
    &\qquad \times \theta(\xi E_P - m_\alpha)\, \theta((1-\xi) E_P - m_\beta)\,.
\end{align}
This decomposition allows us to analyze the divergences associated with the splittings that involve massive partons:
\begin{itemize}
\item 
$g\to q+\Bar{q}$: 
In this splitting, the quarks have masses $m_I=m_J=m_q$. The integral is not divergent as the possible collinear divergences are regulated by the quark masses, while the soft divergences do not arise since the $\theta$-functions restrict the $\xi$-integral to be finite.
\item 
$q\to q+g$: 
In this case, we have one massive quark with mass $m_I=m_P=m_q$. Here, we find only a quasi-collinear enhancement and no divergences outside the soft limit. The details, including the interesting correspondence to the massless scenario, are discussed in Appendix~\ref{appendix:quasi_collinear_enhancement}.
\end{itemize}
We conclude that we only need to consider massless partons to extract the collinear divergences from time-like splittings. Using the results from~\cite{Becher:2023mtx}, we obtain the pole terms 
\begin{align}\label{eq:divergence_hard_function_collinear_splitting_timelike}
    &\langle \boldsymbol{\mathcal{H}}_{m+1}(\argsets{n}{v},\argset{m},\varepsilon) \otimes_{\!\int} \boldsymbol{\mathcal{W}}_{m+1}(\argsets{n}{v},\varepsilon) \rangle \nonumber \\
    &\qquad\supset\frac{\als}{4\pi} \int\! d\mathcal{E}_m \sum_{i\neq 1,2}\left[ C_i \gamma_0 \left(\frac{1}{2\varepsilon^2}+ \frac{1}{\varepsilon} \ln\!\left(\frac{\mu}{2 E_i}\right)\right) - \frac{\gamma_0^i}{\varepsilon} \right] \nonumber \\
    &\;\;\qquad \qquad \times \langle \widetilde{\boldsymbol{\mathcal{H}}}_{m}(\argsets{\hat{n}}{\hat{v}},\argset{\hat{m}},\mu) \otimes_{\!\int}  \boldsymbol{\mathcal{W}}_{m}(\argsets{\hat{n}}{\hat{v}},\mu)  \rangle + \mathcal{O}(\varepsilon^0)\,,
\end{align}
where we have summed over all massless parent partons $P$ in the final state.

Having determined all divergences of the hard functions, we can now write down the anomalous dimension, which we present in the next section.

\subsection{Anomalous dimension of the hard functions}
\label{subsec:anomalous-dimension-hard-function}

To resum the large logarithms arising from scale separation, one needs to understand how the hard functions $\boldsymbol{\mathcal{H}}_m$ evolve with the renormalization scale $\mu$. This evolution is governed by the RG equation
\begin{align}\label{eq:RGE_hard_function}
    \frac{d}{d\ln\mu} \boldsymbol{\mathcal{H}}_m(\argsets{n}{v}, \argset{m}, s, \mu) = - \!\sum^{m}_{l=2+M}\! \boldsymbol{\mathcal{H}}_l(\argsets{n}{v}, \argset{m}, s, \mu) * \boldsymbol{\Gamma}_{lm}^{\mathcal{H}}(\argsets{n}{v}, s, \mu)\,,
\end{align}
where the asterisk denotes a convolution over the momentum fractions for each initial-state parton~\cite{Becher:2023mtx}. As previously discussed, real emissions $\boldsymbol{R}_l$ change the multiplicity of the hard function, whereas virtual corrections $\boldsymbol{V}_l$ leave it unchanged. As a result, the anomalous dimension matrix $\boldsymbol{\Gamma}^{\mathcal{H}}$ takes the form of an upper triangular matrix in multiplicity space,
\begin{align}\label{eq:matrix_form_GammaH}
    \boldsymbol{\Gamma}^{\mathcal{H}}(\argsets{n}{v}, s, \mu) = \frac{\als}{4\pi} \begin{pmatrix}
        \boldsymbol{V}_{2+M} & \boldsymbol{R}_{2+M} & 0 & 0 & \cdots \\
        0 & \boldsymbol{V}_{2+M+1} & \boldsymbol{R}_{2+M+1} & 0 & \cdots \\
        0 & 0 & \boldsymbol{V}_{2+M+2} & \boldsymbol{R}_{2+M+2} & \cdots \\
        \vdots & \vdots & \vdots & \vdots & \ddots
    \end{pmatrix} + \mathcal{O}\left(\alpha^2_s\right).
\end{align}
At one-loop order, the full anomalous dimension can be decomposed into soft and purely collinear parts
\begin{align}\label{eq:split_GammaH_soft_collinear}
    \boldsymbol{\Gamma}^{\mathcal{H}}(\xi_1, \xi_2) = \delta(1-\xi_1)\delta(1-\xi_2) \boldsymbol{\Gamma}^S + \delta(1-\xi_2)\boldsymbol{\Gamma}_1^C(\xi_1) + \delta(1-\xi_1)\boldsymbol{\Gamma}_2^C(\xi_2)\,.
\end{align}
Here, the $\delta$-functions in the first term arise because the momenta of the soft partons are negligible. The purely collinear $\boldsymbol{\Gamma}^C$ leads only to subleading corrections~\cite{Becher:2021zkk} that we will neglect in our analysis. The soft contribution can be further decomposed into \cite{Boer:2024hzh}
\begin{align}
     \boldsymbol{\Gamma}^S =
     \frac{\als}{4\pi}\,\overline{\boldsymbol{\Gamma}}+
     \gcusp\left(\als\right)
     \left[
      \boldsymbol{\Gamma}^c \ln\!\left(\frac{\mu^2}{\mu_h^2}\right) 
    + \boldsymbol{V}^G
    + \boldsymbol{V}^\text{Coul}
     \right]
    + \mathcal{O}(\alpha^2_s)\,. \label{eq:split_GammaS}
\end{align}
The quantity $\overline{\boldsymbol{\Gamma}}$ accounts for the purely soft corrections. We separate the imaginary soft contribution into the usual Glauber term $\boldsymbol{V}^G$ and a new Coulomb-like term $\boldsymbol{V}^{\text{Coul}}$. The soft\,+\,collinear terms are contained in the cusp contribution $\boldsymbol{\Gamma}^c$.

Collecting all contributions to the soft anomalous dimension in the presence of massive final states, we obtain
\begin{align}
    \overline{\boldsymbol{\Gamma}} =& \frac{1}{2} \gamma_0 \sum_{\alpha, \beta}(\boldsymbol{T}_{\alpha, L} \cdot \boldsymbol{T}_{\beta, L} + \boldsymbol{T}_{\alpha, R} \cdot \boldsymbol{T}_{\beta, R}) \int \frac{d^2\Omega_k}{4\pi} \overline{W}_{\alpha\beta}^k - 4 \sum_{\alpha, \beta} \theta_{\mathrm{hard}}(n_k) \overline{W}_{\alpha\beta}^k \boldsymbol{T}_{\alpha,L} \circ \boldsymbol{T}_{\beta,R}\,, \nonumber
    \\
    \boldsymbol{\Gamma}^c =& \sum_{i=1,2} \left[C_i \, \boldsymbol{1} - \delta(n_i - n_k)\, \boldsymbol{T}_{i,L} \circ \boldsymbol{T}_{i,R}\right], \nonumber
    \\
    \boldsymbol{V}^G =& -2\pi i \left(\boldsymbol{T}_{1,L} \cdot \boldsymbol{T}_{2,L} - \boldsymbol{T}_{1,R} \cdot \boldsymbol{T}_{2,R} \right),\nonumber
    \\
    \boldsymbol{V}^\mathrm{Coul} =& -\frac{1}{2}\pi i \sum_{(IJ)} \left(\boldsymbol{T}_{I,L} \cdot \boldsymbol{T}_{J,L} - \boldsymbol{T}_{I,R} \cdot \boldsymbol{T}_{J,R}\right)v_{IJ}\,. 
     \label{eq:final_expression_Gammabar}
\end{align}

\section{New sources of super-leading logarithms}
\label{sec:new-sources-of-SLLs}

The hard functions can be evolved from the hard scale set by the partonic center-of-mass energy, $\mu_h \sim \sqrt{\hat{s}}$, down to the scale $\mu_s$ relevant for the low-energy matrix element, by solving the RG equation \eqref{eq:RGE_hard_function}. The solution is given by
\begin{align}
    \boldsymbol{\mathcal H}(\mu_s) = \boldsymbol{\mathcal H}(\mu_h) \ast \boldsymbol{U}(\mu_h, \mu_s)\,,
\end{align}
where the corresponding evolution operator is represented by a path-ordered exponential, 
\begin{align}\label{eq:evolution_operator_U}
    \boldsymbol{U}(\mu_h, \mu_s) = \mathbf{P} \exp\left[\int_{\mu_s}^{\mu_h}\frac{d\mu'}{\mu'} \boldsymbol{\Gamma}^{\mathcal{H}}(\mu')\right].
\end{align}
Writing the exponential as a power series, the scale-evolved hard functions can be expressed as
\begin{align}\label{eq:hard_function_evolved_to_soft_scale}
    \boldsymbol{\mathcal{H}}(\mu_s) = \boldsymbol{\mathcal{H}}(\mu_h) &+ \int_{\mu_s}^{\mu_h}\frac{d\mu_1}{\mu_1} \boldsymbol{\mathcal{H}}(\mu_h) \ast \boldsymbol{\Gamma}^{\mathcal{H}}(\mu_1) \nonumber \\
    &+ \int_{\mu_s}^{\mu_h}\frac{d\mu_1}{\mu_1} \int_{\mu_s}^{\mu_1}\frac{d\mu_2}{\mu_2} \boldsymbol{\mathcal{H}}(\mu_h) \ast \boldsymbol{\Gamma}^{\mathcal{H}}(\mu_1) \ast \boldsymbol{\Gamma}^{\mathcal{H}}(\mu_2) + \dots
\end{align}
with the ordering $\mu_h>\mu_1> \mu_2 > \dots>\mu_s$. Only specific combinations of terms in $\boldsymbol{\Gamma}^{\mathcal{H}}$ generate SLLs, as we will discuss in detail in the following.

\subsection{Appearance of additional color structures}
\label{subsec:add-color-struc}

Using the results from Section~\ref{subsec:anomalous-dimension-hard-function}, it follows that different parts of the anomalous dimension act on the hard functions. The SLLs arise from non-vanishing terms with the maximal number of insertions of $\boldsymbol{\Gamma}^c$ at a given order in $\alpha_s$. Certain properties of the parts of $\boldsymbol{\Gamma}^{S}$ simplify our discussion. Besides the known commutation relation encoding color coherence \cite{Becher:2021zkk},
\begin{align}\label{eq:identities_anomalous_dimension_commutator_1}
    \left[\boldsymbol{\Gamma}^c, \overline{\boldsymbol{\Gamma}}\right] = 0\,,
\end{align}
we find relations involving the Coulomb term,
\begin{align}\label{eq:identities_anomalous_dimension_commutator_2}
    \left[\boldsymbol{V}^G, \boldsymbol{V}^\text{Coul}\right] = 0 \qquad \text{and}  \qquad
    \left[\boldsymbol{\Gamma}^c, \boldsymbol{V}^\text{Coul}\right] = 0\,,
\end{align}
because the operators act on different partons. Moreover, for generic hard functions, we have~\cite{Becher:2021zkk}
\begin{align}\label{eq:identities_anomalous_dimension_trace}
    \langle \boldsymbol{\mathcal{H}} \boldsymbol{\Gamma}^c \otimes_{\!\int}\!\boldsymbol{1} \rangle = 0\,, \qquad
    \langle \boldsymbol{\mathcal{H}} \boldsymbol{V}^G \otimes_{\!\int}\!\boldsymbol{1} \rangle = 0\,, \qquad
    \langle \boldsymbol{\mathcal{H}} \boldsymbol{V}^\text{Coul} \otimes_{\!\int}\!\boldsymbol{1} \rangle = 0
\end{align}
by the cyclicity of the trace.

In addition to the known SLLs~\cite{Becher:2021zkk}, new sources of SLLs arise from substituting insertions of $\boldsymbol{V}^G$ by $\boldsymbol{V}^\text{Coul}$.
Because the latter commutes with $\boldsymbol{\Gamma}^c$, however, insertions of $\boldsymbol{\Gamma}^c$ give non-vanishing contributions only if at least one Glauber operator is present. Therefore, to obtain a SLL and a real contribution to the cross section, we can only substitute one Glauber operator with a Coulomb phase. This leads to the possible color structures
\begin{align}\label{eq:Crn_Coulomb}
    C_{rn} & =  \langle \boldsymbol{\mathcal{H}}_{2\to M} (\boldsymbol{\Gamma}^c)^r \boldsymbol{V}^G (\boldsymbol{\Gamma}^c)^{n-r} \boldsymbol{V}^G \overline{\boldsymbol{\Gamma}} \otimes_{\!\int}\!\boldsymbol{1}\rangle\,, \nonumber \\
    C_{rn}^{\Coul,a} & =  \langle \boldsymbol{\mathcal{H}}_{2\to M} (\boldsymbol{\Gamma}^c)^r \boldsymbol{V}^\Coul (\boldsymbol{\Gamma}^c)^{n-r} \boldsymbol{V}^G \overline{\boldsymbol{\Gamma}} \otimes_{\!\int}\!\boldsymbol{1}\rangle\,,  \nonumber \\
    C_{n}^{\Coul,b} & =  \langle \boldsymbol{\mathcal{H}}_{2\to M} (\boldsymbol{\Gamma}^c)^n  \boldsymbol{V}^G \boldsymbol{V}^\Coul  \overline{\boldsymbol{\Gamma}} \otimes_{\!\int}\!\boldsymbol{1}\rangle\,,
\end{align}
where $0\leq r \leq n$ and $\boldsymbol{\mathcal{H}}_{2\to M}$ denotes the Born-level hard function. By virtue of the commutation relations~\eqref{eq:identities_anomalous_dimension_commutator_2}, the last two terms yield only one new color structure
\begin{align}\label{eq:Cn_Coulomb}
    C_{n}^\text{Coul} &\equiv C_{rn}^{\Coul,a} = C_{n}^{\Coul,b}
    = \langle \boldsymbol{\mathcal{H}}_{2\to M} \boldsymbol{V}^\text{Coul} (\boldsymbol{\Gamma}^c)^n \boldsymbol{V}^G \overline{\boldsymbol{\Gamma}} \otimes_{\!\int}\!\boldsymbol{1}\rangle\,.
\end{align}
Note that while the order of the operators does not affect the color traces, it does affect the corresponding scale integrals. We will discuss this in Section~\ref{subsec:scale-int}.

As found in~\cite{Becher:2021zkk}, the terms with $n=0$ contributing at $\mathcal{O}(\alpha^3_s)$, even though not strictly yielding SLLs, can still be numerically relevant. Therefore, we include them in our discussion and consider the additional structure involving two Coulomb insertions,
\begin{align}\label{eq:C_2Coulomb}
    C^{\Coul^2} & =  \langle \boldsymbol{\mathcal{H}}_{2\to M} \boldsymbol{V}^\Coul \boldsymbol{V}^\Coul \overline{\boldsymbol{\Gamma}} \otimes_{\!\int}\!\boldsymbol{1}\rangle\,,
\end{align}
which would vanish with any insertion of $\boldsymbol{\Gamma}^c$.

\subsection{Evaluation of the color structures}
\label{subsec:eval-color-struc}

We now demonstrate the evaluation of the new Coulomb-induced contributions arising from \eqref{eq:Cn_Coulomb} and \eqref{eq:C_2Coulomb}. We will refer to the corresponding terms as Coulomb SLLs. For the pure Glauber contributions (Glauber SLLs), we refer the reader to \cite{Becher:2021zkk}.

\subsubsection{Single Coulomb insertion}

We start by considering the color structure $C_{n}^\text{Coul}$ defined in \eqref{eq:Cn_Coulomb}. First, we use that for an arbitrary hard function $\boldsymbol{\mathcal{H}}$,\footnote{We will repeatedly use the symbol $\boldsymbol{\mathcal{H}}$ below to indicate any hard functions. It is understood to denote different combinations of $\boldsymbol{\mathcal{H}}_{2\to M}$ together with operators from the anomalous dimension.}
\begin{align}\label{eq:simplifying_color_H_VG_Gammabar}
    \langle \boldsymbol{\mathcal{H}} \boldsymbol{V}^G \overline{\boldsymbol{\Gamma}} \otimes_{\!\int}\!\boldsymbol{1} \rangle = 16 \pi i \langle \boldsymbol{\mathcal{H}}  \boldsymbol{X}_1 \otimes_{\!\int}\!\boldsymbol{1} \rangle\,,
\end{align}
which is analogous to the massless case considered in \cite{Becher:2023mtx}.\footnote{We follow the notation of \cite{Boer:2024hzh}, where factors of $\gamma_0=4$ have been moved from the color traces $C_{rn}$ to the scale integrals $I_{rn}$ defined in \eqref{eq:scale_integrals_Irn_general_expression}.} In this expression, the operator reads
\begin{align}\label{eq:X1 and Jbeta}
    \boldsymbol{X}_1 = i f^{abc} \sum_{\beta \neq 1,2} J_{\beta}   \boldsymbol{T}_1^a \boldsymbol{T}_2^b \boldsymbol{T}_\beta^c 
\qquad \text{with} \qquad
    J_\beta = \int \frac{d^2\Omega_k}{4\pi} \left(W_{1\beta}^k - W_{2\beta}^k \right) \theta_\text{veto}(n_k)\,,
\end{align}
where we have used $\theta_\text{veto}=1-\theta_\text{hard}$, and $f^{abc}$ are the totally antisymmetric structure constants. This shows that the massive monopoles from $\overline{\boldsymbol{\Gamma}}$ no longer contribute after the application of $\boldsymbol{V}^G$. It can then be shown that $\boldsymbol{\Gamma}^c$ acts in a simple way on $\boldsymbol{X}_1$ \cite{Becher:2023mtx}, leading to
\begin{align}\label{eq:simplifying_color_H_Gammacuspn_VG_Gammabar}
    \langle \boldsymbol{\mathcal{H}} (\boldsymbol{\Gamma}^c)^n \boldsymbol{V}^{G} \overline{\boldsymbol{\Gamma}} \otimes_{\!\int}\!\boldsymbol{1} \rangle = 16 \pi i (N_c)^n \langle \boldsymbol{\mathcal{H}}  \boldsymbol{X}_1 \otimes_{\!\int}\!\boldsymbol{1} \rangle\,.
\end{align}
The next step is to determine the action of $\boldsymbol{V}^\text{Coul}$ on $\boldsymbol{X}_1$. Explicitly, it reads
\begin{align}\label{eq:simplifying_color_H_VCoul_X1_first_step}
    \langle \boldsymbol{\mathcal{H}} \boldsymbol{V}^\text{Coul} \boldsymbol{X}_1 \otimes_{\!\int}\!\boldsymbol{1} \rangle &= \frac{\pi}{2} f^{abc} \sum_{(IJ)}  v_{IJ} \sideset{}{'}\sum_{\beta\neq 1,2} \langle \boldsymbol{\mathcal{H}} J_\beta \left(\boldsymbol{T}_{I,L} \cdot \boldsymbol{T}_{J,L} - \boldsymbol{T}_{I,R} \cdot \boldsymbol{T}_{J,R}\right) \boldsymbol{T}_1^a \boldsymbol{T}_2^b \boldsymbol{T}_\beta^c \otimes_{\!\int}\!\boldsymbol{1} \rangle\,,
\end{align}
where the prime on the sum indicates that collinear gluons emitted from $\boldsymbol{\Gamma}^c$ are not included.  Note that for the massive case, the angular integrals $J_I$ depend on the parton energy and therefore cannot be factored out of the bracket, which includes the integration over these energies. The color structure can be rewritten as 
\begin{align}\label{eq:simplifying_color_H_VCoul_X1_second_step}
    \langle \boldsymbol{\mathcal{H}} \left(\boldsymbol{T}_{I,L} \cdot \boldsymbol{T}_{J,L} - \boldsymbol{T}_{I,R} \cdot \boldsymbol{T}_{J,R}\right) \boldsymbol{T}_1^a \boldsymbol{T}_2^b \boldsymbol{T}_\beta^c \otimes_{\!\int}\!\boldsymbol{1} \rangle 
    = \langle \boldsymbol{\mathcal{H}} \boldsymbol{T}_1^a \boldsymbol{T}_2^b ( \boldsymbol{T}_\beta^c  \boldsymbol{T}_{I}^d  \boldsymbol{T}_{J}^d   - \boldsymbol{T}_{I}^d  \boldsymbol{T}_{J}^d  \boldsymbol{T}_\beta^c ) \otimes_{\!\int}\!\boldsymbol{1} \rangle\,,
\end{align}
where we used that $\boldsymbol{T}_I$ and $\boldsymbol{T}_J$ commute with $\boldsymbol{T}_{1,2}$. We now consider the different terms in the sum over $\beta$. For $\beta \neq I,J$, the color generators commute and vanish, such that we can set $\beta$ to $I$ and $J$. After simplifying the commutators through $\left[\boldsymbol{T}_\alpha^a, \boldsymbol{T}_\beta^b\right] = i f^{abc} \boldsymbol{T}_\alpha^c \delta_{\alpha \beta}$, we get
\begin{align}\label{eq:simplifying_color_H_VCoul_X1_fourth_step}
    \langle \boldsymbol{\mathcal{H}} V^\text{Coul} \boldsymbol{X}_1 \otimes_{\!\int}\!\boldsymbol{1} \rangle = \frac{i\pi}{2} f^{abc} f^{cde}   \sum_{(IJ)}  v_{IJ} \, \langle \boldsymbol{\mathcal{H}} (J_I - J_J)\boldsymbol{T}_1^a \boldsymbol{T}_2^b \boldsymbol{T}_I^e \boldsymbol{T}_J^d \otimes_{\!\int}\!\boldsymbol{1}\rangle \,.
\end{align}
The term $(J_I-J_J)$ simplifies to $2J_I$. After swapping the indices $c$ and $e$, we define the operator 
\begin{align}\label{eq:XCoul}
    \boldsymbol{X}^\text{Coul} &\equiv - f^{abe}  f^{cde}   \sum_{(IJ)}  v_{IJ} J_I  \boldsymbol{T}_1^a \boldsymbol{T}_2^b \boldsymbol{T}_I^c \boldsymbol{T}_J^d \,,
\end{align}
such that
\begin{align}\label{eq:simplifying_color_H_VCoul_X1_result}
    \langle \boldsymbol{\mathcal{H}}_{2\to M} \boldsymbol{V}^\text{Coul} \boldsymbol{X}_1 \otimes_{\!\int}\!\boldsymbol{1} \rangle = i \pi \langle \boldsymbol{\mathcal{H}} \boldsymbol{X}^\text{Coul} \otimes_{\!\int}\!\boldsymbol{1} \rangle \,.
\end{align}
This gives us the final expression
\begin{align}\label{eq:simplifying_color_CnCoul_result}
    C_n^\text{Coul} =\langle \boldsymbol{\mathcal{H}}_{2\to M} \boldsymbol{V}^\text{Coul} (\boldsymbol{\Gamma}^c)^n \boldsymbol{V}^{G} \overline{\boldsymbol{\Gamma}} \otimes_{\!\int}\!\boldsymbol{1} \rangle = - 16 \pi^2 (N_c)^n \langle \boldsymbol{\mathcal{H}}_{2\to M} \boldsymbol{X}^\text{Coul} \otimes_{\!\int}\!\boldsymbol{1} \rangle\,.
\end{align}

\subsubsection{Double Coulomb insertion}

To evaluate the double Coulomb insertion $C^{\Coul^2}$ in~\eqref{eq:C_2Coulomb}, we start from
\begin{align}\label{eq:trace_H_GammaBar}
    \langle \boldsymbol{\mathcal{H}} \overline{\boldsymbol{\Gamma}} \otimes_{\!\int}\!\boldsymbol{1} \rangle = 4 \sum_{\alpha,\beta} \langle  \boldsymbol{\mathcal{H}}  \tope_\alpha \cdot \tope_\beta \int \frac{d^2\Omega_{k_0}}{4\pi} W_{\alpha\beta}^{k_0} \theta_\text{veto}(n_{k_0}) \otimes_{\!\int}\!\boldsymbol{1} \rangle \,.
\end{align}
Inserting one Coulomb operator gives us
\begin{align}\label{eq:trace_H_VCoul_GammaBar}
    \langle \boldsymbol{\mathcal{H}} \boldsymbol{V}^\Coul \overline{\boldsymbol{\Gamma}} \otimes_{\!\int}\!\boldsymbol{1} \rangle 
    =-4\pi f^{abc} \sum_{(IJ)}\sum_{\alpha \neq I,J} v_{IJ} 
    \langle \boldsymbol{\mathcal{H}} \tilde{J}_\alpha^{IJ}\tope_I^a \tope_J^b \tope_\alpha^c \otimes_{\!\int}\!\boldsymbol{1} \rangle\,,
\end{align}
where the angular integrals are defined as
\begin{align}\label{eq:massive_angular_integral}
    \tilde{J}_\alpha^{IJ} = \int \frac{d^2\Omega_k}{4\pi} \left(W_{I\alpha}^k - W_{J\alpha}^k \right) \theta_\text{veto}(n_k)\,,
\end{align}
in analogy to \eqref{eq:X1 and Jbeta}. Note the strong similarity to the single Glauber insertion given in \eqref{eq:simplifying_color_H_VG_Gammabar}. For the second Coulomb insertion, we now have to evaluate
\begin{align}\label{eq:trace_H_VCoul_VCoul_GammaBar}
    \langle \boldsymbol{\mathcal{H}} \boldsymbol{V}^\Coul \boldsymbol{V}^\Coul \overline{\boldsymbol{\Gamma}} \otimes_{\!\int}\!\boldsymbol{1} \rangle &= 2\pi^2 i f^{abc} \sum_{(IJ)}\sum_{\alpha \neq I,J} v_{IJ} \nonumber \\
    &\qquad \times\sum_{(MN)} v_{MN} 
    \langle \boldsymbol{\mathcal{H}}  
    \tilde{J}_\alpha^{IJ} 
    (\tope_{M,L}\cdot\tope_{N,L}-\tope_{N,R}\cdot\tope_{M,R}) \tope_I^a \tope_J^b \tope_\alpha^c \otimes_{\!\int}\!\boldsymbol{1} \rangle\,.
\end{align}
In order to do so, we split the sum $\sum_{(MN)}$ into the cases where, i) $M=I$ and $N=J$ or vice versa (the second case yields a factor of 2), ii)  $M=I,J$ and $N\neq I,J$ or vice versa (again factor of 2), and iii)  $M,N\neq I,J$. After some color algebra, we arrive at
\begin{align}\label{eq:C2Coul_trace_evaluated}
    C^{\Coul^2}&=\langle \boldsymbol{\mathcal{H}} \boldsymbol{V}^\Coul \boldsymbol{V}^\Coul \overline{\boldsymbol{\Gamma}} \otimes_{\!\int}\!\boldsymbol{1} \rangle \nonumber \\
    &= -4\pi^2 f^{abe} f^{cde} \sum_{(IJ)}v_{IJ} \bigg\{\sum_{\alpha \neq I,J} v_{IJ} \, 
    \langle \boldsymbol{\mathcal{H}}\tilde{J}_\alpha^{IJ} \tope_I^c \{\tope_J^b,\tope_J^d\}\tope_\alpha^a \otimes_{\!\int}\!\boldsymbol{1} \rangle \nonumber \\
    &\qquad 
    +2 \sum_{\alpha \neq I,J} \sum_{N\neq I,J} v_{IN}  \,
    \langle  \boldsymbol{\mathcal{H}} \tilde{J}_\alpha^{IJ}\tope_I^c \tope_J^b \tope_\alpha^a \tope_N^d \otimes_{\!\int}\!\boldsymbol{1} \rangle \nonumber \\
    &\qquad 
    -2 \sum_{N \neq I,J} v_{IN} \, 
    \langle \boldsymbol{\mathcal{H}} \tilde{J}_N^{IJ}\tope_I^d \tope_I^a \tope_J^b \tope_N^c \otimes_{\!\int}\!\boldsymbol{1} \rangle \nonumber \\
    &\qquad 
    - \! \sum_{(MN) \neq I,J} \! v_{MN} \,  
    \langle \boldsymbol{\mathcal{H}} \tilde{J}_M^{IJ}\tope_I^a \tope_J^b \tope_M^c \tope_N^d \otimes_{\!\int}\!\boldsymbol{1} \rangle \bigg\} \nonumber \\
    &\equiv -4\pi^2 \langle \boldsymbol{\mathcal{H}} \boldsymbol{X}^{\Coul^2} \otimes_{\!\int}\!\boldsymbol{1} \rangle.
\end{align}
The first line, which corresponds to i), requires only one pair of massive final states. This is the contribution relevant for $2\to t\Bar{t}$ scattering, which we will discuss in detail later. Contrary to that, the second and third line, corresponding to ii), involve at least three heavy quarks, while the fourth line -- from case iii) -- requires at least four heavy quarks in the final state. For processes of the type $2\to t\bar{t}+\text{light}$, these are therefore not relevant. The last line defines $\boldsymbol{X}^{\Coul^2}$, the operator in color space that encodes the double Coulomb insertion.

\subsection{Performing the scale integrals}
\label{subsec:scale-int}

While the ordering of the operators is not relevant for the color traces, we need to take it into account when evaluating the scale integrals in relation \eqref{eq:hard_function_evolved_to_soft_scale}. For the case of $C_{rn}^{\text{Coul},a}$ in \eqref{eq:Crn_Coulomb}, the ordering reads
\begin{align}\label{eq:scale_integrals_Irn_general_expression}
    I_{rn}(\mu_h, \mu_s) =& \int_{\mu_s}^{\mu_h} \!\frac{d\mu_1}{\mu_1} \gcuspals{\mu_1} \ln\!\left(\frac{\mu_1^2}{\mu_h^2}\right) \! \dots \!
    \int_{\mu_s}^{\mu_{r-1}} \!\frac{d\mu_r}{\mu_r} \gcuspals{\mu_r} \ln\!\left(\frac{\mu_r^2}{\mu_h^2}\right) \nonumber \\
     \times &\int_{\mu_s}^{\mu_r} \!\frac{d\mu_{r+1}}{\mu_{r+1}} \gcuspals{\mu_{r+1}} \nonumber \\
     \times &\int_{\mu_s}^{\mu_{r+1}} \!\frac{d\mu_{r+2}}{\mu_{r+2}} \gcuspals{\mu_{r+2}} \ln\!\left(\hspace{-0.25pt}\frac{\mu_{r+2}^2}{\mu_h^2}\hspace{-0.25pt}\right) \! \dots \!
    \int_{\mu_s}^{\mu_n} \!\frac{d\mu_{n+1}}{\mu_{n+1}} \gcuspals{\mu_{n+1}} \ln\!\left(\hspace{-0.25pt}\frac{\mu_{n+1}^2}{\mu_h^2}\hspace{-0.25pt}\right) \nonumber \\
    \times  &\int_{\mu_s}^{\mu_{n+1}} \!\frac{d\mu_{n+2}}{\mu_{n+2}} \gcuspals{\mu_{n+2}} 
    \int_{\mu_s}^{\mu_{n+2}} \!\frac{d\mu_{n+3}}{\mu_{n+3}} \frac{\als(\mu_{n+3})}{4\pi}\,.
\end{align}
For $C_{n}^{\text{Coul},b}$ in \eqref{eq:Crn_Coulomb}, we obtain $I_{nn}$ (setting $r=n$ in the equation above). In evaluating the scale integrals, we account for the top-quark threshold since the top-quark mass lies between $\mu_h$ and $\mu_s$. For readability, we omit the threshold dependence in the results below, but do include it in our numerical analysis.

\subsubsection{Fixed-coupling approximation}

We can simplify the evaluation of the scale integrals in the approximation of a fixed coupling $\als(\Bar{\mu})$ at an intermediate scale, treating the running of the coupling as a subleading effect. This enables us to write the resummed expression in a closed form.
We have \cite{Becher:2021zkk}
\begin{align}\label{eq:scale_integrals_Irn_fixed_coupling}
    I_{rn}(\mu_h, \mu_s)
    = \left(\frac{\als(\Bar{\mu})}{4\pi}\right)^{n+3} \gamma_0^{n+2} \frac{(-4)^n n!}{(2n+3)!} \frac{(2r)!}{4^r(r!)^2} \ln^{2n+3}\!\left(\frac{\mu_h}{\mu_s}\right).
\end{align}
The partonic cross-section with the resummed tower of Coulomb SLLs for a single Coulomb insertion is then given by
\begin{align}\label{eq:sigma_SLL_Coul_fixed_coupling_expressed_in_Irn}
    \hat{\sigma}_{2\to M}^{\text{SLL,Coul}} &=\sum_{n=0}^\infty \left[ I_{nn}(\mu_h,\mu_s) C_n^{\text{Coul},b} + \sum_{r=0}^n I_{rn}(\mu_h, \mu_s) C_{rn}^{\text{Coul},a}\right] \nonumber \\
    &= \sum_{n=0}^\infty C_n^\text{Coul}\left[I_{nn}(\mu_h,\mu_s)+\sum_{r=0}^n I_{rn}(\mu_h, \mu_s)\right].
\end{align}
We now use $w=N_c(\als(\Bar{\mu})/\pi) \ln^2(\mu_h/\mu_s)$ and perform the sums over $r$ and $n$. The result for the partonic cross section in the fixed-coupling approximation then reads
\begin{align}\label{eq:sigma_SLL_1Coul_fixed_coupling_result}
    \hat{\sigma}_{2\to M}^{\text{SLL,Coul}} &= - 16 \pi^2 \langle \boldsymbol{\mathcal{H}}_{2\to M}(\mu_h) \boldsymbol{X}^{\text{Coul}} \otimes_{\!\int}\!\boldsymbol{1} \rangle \times 4\left(\frac{\als(\bar{\mu})}{4\pi}\right)^3 \ln^3\!\left(\frac{\mu_h}{\mu_s}\right) \nonumber\\
   &\qquad \times \left[\frac{e^{-w}}{w} + \frac{\sqrt{\pi}(-1+2w) 
   \erf(\sqrt{w})}{2w^{3/2}}\right].
\end{align}
Analogously, for the double Coulomb insertion, where only $I_{00}(\mu_h, \mu_s)$ contributes, we obtain
\begin{align}\label{eq:sigma_SLL_2Coul_fixed_coupling_result}
    \hat{\sigma}_{2\to M}^{\text{SLL,2\,Coul}} = - 4 \pi^2 \langle \boldsymbol{\mathcal{H}}_{2\to M}(\mu_h) \boldsymbol{X}^{\text{Coul}^2} \otimes_{\!\int}\!\boldsymbol{1} \rangle \times\frac{8}{3}\left(\frac{\als(\Bar{\mu})}{4\pi}\right)^3 \ln^3\!\left(\frac{\mu_h}{\mu_s}\right).
\end{align}

\subsubsection{RG-improved resummation}

Instead of using the fixed-coupling approximation, we now switch to RG-improved resummation, following \cite{Boer:2024hzh}. For a single Coulomb insertion, the $(n+3)$-scale integrals in $I_{rn}$ can be simplified by introducing the generalized Sudakov operator 
\begin{align}\label{eq:Uc}
    \boldsymbol{U}^c(\mu_i, \mu_j) = \exp\left[\boldsymbol{\Gamma}^c \int_{\mu_j}^{\mu_i} \frac{d\mu}{\mu} \gcuspals{\mu} \ln\!\left(\frac{\mu^2}{\mu_h^2}\right)\right].
\end{align}
With this, the evolution operator producing SLLs for a single Coulomb insertion reads
\begin{align}\label{eq:U_SLL_Coul}
    \boldsymbol{U}_{\text{SLL}}^{\text{Coul}}&(\mu_h, \mu_s) = \int_{\mu_s}^{\mu_h}\frac{d\mu_1}{\mu_1} \int_{\mu_s}^{\mu_1}\frac{d\mu_2}{\mu_2} \int_{\mu_s}^{\mu_2}\frac{d\mu_3}{\mu_3} \boldsymbol{U}^c(\mu_h, \mu_1) \gcuspals{\mu_1} \nonumber\\
    \times& \Bigl[
    \boldsymbol{V}^{\text{Coul}} \boldsymbol{U}^c(\mu_1, \mu_2) \gcuspals{\mu_2} \boldsymbol{V}^G 
    + \boldsymbol{V}^G \gcuspals{\mu_2} \boldsymbol{V}^{\text{Coul}} 
    \Bigr] 
    \frac{\als(\mu_3)}{4\pi} \overline{\boldsymbol{\Gamma}}\,,
\end{align}
where the first term corresponds to the color structure $C_{rn}^{\text{Coul},a}$ and the second term to $C_n^{\text{Coul},b}$. The eigenvalue of $N_c$ for each insertion of $\Gamma^c$ now exponentiates to 
\begin{align}\label{eq:Uc_eigenvalue}
    U^c(1; \mu_i, \mu_j) = \exp\left[ N_c \int_{\mu_j}^{\mu_i} \frac{d\mu}{\mu} \gcuspals{\mu}\ln\!\left(\frac{\mu^2}{\mu_h^2}\right)\right] .
\end{align}
In the first term of \eqref{eq:U_SLL_Coul} the two factors combine, $U^c(1; \mu_h, \mu_1)\, U^c(1; \mu_1, \mu_2) = U^c(1; \mu_h, \mu_2)$, while the second line contains only $U^c(1; \mu_h, \mu_1)$.
In order to obtain the partonic cross-section, we need to evaluate
\begin{align}\label{eq:SLL_Coul_RG_improved_three_scale_integrals}
    &\langle \boldsymbol{\mathcal{H}}_{2\to M}(\mu_h) \boldsymbol{U}_{\text{SLL}}^{\text{Coul}}(\mu_h, \mu_s) \otimes_{\!\int}\!\boldsymbol{1} \rangle  = -16 \pi^2 \langle \boldsymbol{\mathcal{H}}_{2\to M}(\mu_h)  \boldsymbol{X}^\text{Coul} \otimes_{\!\int}\!\boldsymbol{1} \rangle \nonumber \\
    &\qquad  \times  16\int_{\mu_s}^{\mu_h}\frac{d\mu_1}{\mu_1} \int_{\mu_s}^{\mu_1}\frac{d\mu_2}{\mu_2} \int_{\mu_s}^{\mu_2}\frac{d\mu_3}{\mu_3} \frac{\als(\mu_1)}{4\pi} \frac{\als(\mu_2)}{4\pi} \frac{\als(\mu_3)}{4\pi}  \left[U^c(1; \mu_h, \mu_2) + U^c(1; \mu_h, \mu_1)\right].
\end{align}
The integral over $\mu_3$ can be immediately performed, yielding
\begin{align}\label{eq:RGE_als_integrated}
    \int_{\mu_i}^{\mu_j} \frac{d\mu}{\mu} \frac{\als(\mu)}{4\pi} = - \frac{1}{2\beta_0} \ln\!\left(\frac{\als(\mu_j)}{\als(\mu_i)}\right)
\end{align}
at lowest order in $\alpha_s$. In the second term, we can do the same for the integral over $\mu_2$. For the first term, we change the integration order before we can integrate analogously over $\mu_1$. We obtain
\begin{align}\label{eq:SLL_Coul_RG_improved_one_scale_integral}
    &\langle \boldsymbol{\mathcal{H}}_{2\to M}(\mu_h) \boldsymbol{U}_{\text{SLL}}^{\text{Coul}}(\mu_h, \mu_s) \otimes_{\!\int}\!\boldsymbol{1} \rangle  = - 16\pi^2 \langle \boldsymbol{\mathcal{H}}_{2\to M}(\mu_h)  \boldsymbol{X}^{\text{Coul}} \otimes_{\!\int}\!\boldsymbol{1} \rangle \nonumber \\
    &\qquad \times \frac{1}{\beta_0^2}\int_{\mu_s}^{\mu_h}\frac{d\mu}{\mu} \frac{\als(\mu)}{\pi}  U^c(1; \mu_h, \mu) \left[\ln\!\left(\frac{\als(\mu)}{\als(\mu_s)}\right) \ln\!\left(\frac{\als(\mu_h)}{\als(\mu)}\right) + \frac{1}{2} \ln^2\!\left(\frac{\als(\mu)}{\als(\mu_s)}\right) \right].
\end{align}
This leaves us with a single scale integral that has to be performed numerically. We introduce the coupling ratio $x = \als(\mu) / \als(\mu_h)$, in which the Sudakov factor takes the form \cite{Neubert:2004dd}
\begin{align}\label{eq:Uc_eigenvalue_in_x}
    U^c(1;\mu_h,\mu) = \exp\Bigg\{\frac{\gamma_0 N_c}{2\beta_0^2} \bigg[&\frac{4\pi}{\als(\mu_h)}\left(1-\frac{1}{x}-\ln x\right) \nonumber \\
    &+ \left(\frac{\gamma_1}{\gamma_0}-\frac{\beta_1}{\beta_0}\right) \left(1-x+\ln x\right) + \frac{\beta_1}{2\beta_0}\ln^2 x\bigg]\Bigg\}\,.
\end{align}
Here, $\gamma_0$ and $\gamma_1$ are the one- and two-loop coefficients of the light-like cusp anomalous dimension given in \eqref{eq:gammacoefs}, while $\beta_0$ and $\beta_1$ are the first two coefficients of the QCD $\beta$-function. 

In terms of $x$ and $x_s = \als(\mu_s) / \als(\mu_h)$, the full expression for one Coulomb insertion can be written as
\begin{align}\label{eq:SLL_Coul_RG_improved_one_scale_integral_in_x}
    \langle \boldsymbol{\mathcal{H}}_{2\to M}(\mu_h) \boldsymbol{U}_{\text{SLL}}^\Coul(\mu_h, \mu_s) \otimes_{\!\int}\!\boldsymbol{1} \rangle  &= -16\pi^2 \langle \boldsymbol{\mathcal{H}}_{2\to M}(\mu_h)  \boldsymbol{X}^\Coul \otimes_{\!\int}\!\boldsymbol{1} \rangle \nonumber \\
    &\qquad \times  \frac{1}{\beta_0^3}\int_{1}^{x_s}\frac{dx}{x}   U^c(1; \mu_h, \mu) \left(\ln^2 x_s-\ln^2 x \right) .
\end{align}
For the double Coulomb insertion, the evolution operator reads
\begin{align}\label{eq:USLL2Coul}
    \boldsymbol{U}_{\text{SLL}}^{\Coul^2}(\mu_h, \mu_s) =\! \int_{\mu_s}^{\mu_h}\frac{d\mu_1}{\mu_1} \int_{\mu_s}^{\mu_1}\frac{d\mu_2}{\mu_2} \int_{\mu_s}^{\mu_2}\frac{d\mu_3}{\mu_3} \gcuspals{\mu_1} \boldsymbol{V}^\Coul  \gcuspals{\mu_2} \boldsymbol{V}^\Coul \frac{\als(\mu_3)}{4\pi} \overline{\boldsymbol{\Gamma}}.
\end{align}
Evaluating all scale integrals straightforwardly, we obtain
\begin{align}\label{eq:trace_H_USLL2Coul}
    &\langle \boldsymbol{\mathcal{H}}_{2\to M}(\mu_h) \boldsymbol{U}_{\text{SLL}}^{\Coul^2}(\mu_h, \mu_s) \otimes_{\!\int}\!\boldsymbol{1} \rangle  = - 4\pi^2 \langle \boldsymbol{\mathcal{H}}_{2\to M}(\mu_h)  \boldsymbol{X}^{\Coul^2} \otimes_{\!\int}\!\boldsymbol{1} \rangle \cdot \frac{1}{3\beta_0^3} \ln^3 x_s.
\end{align}
This is consistent with the result obtained in the fixed-coupling approximation.

\subsection{Evaluation of the angular integrals}
\label{subsec:angular-integrals}

The color structure $\boldsymbol{X}^{\Coul}$ contains the massive angular integrals $J_I$ defined in \eqref{eq:X1 and Jbeta}. Using the parametrization of the ATLAS experiment \cite{ATLAS:2012al}, these integrals depend on jet pseudorapidities
$\eta_I=\frac{1}{2} \ln\!\left(\frac{1+\cos \theta_I}{1-\cos \theta_I}\right) = \artanh (\cos\theta_I)$ and read
\begin{align} 
    J_I = \frac{1}{2\pi} \int_0^{2\pi} d\phi_k \int_{\eta_a}^{\eta_b} d\eta_k \frac{\sinh \eta_k\cosh\eta_I - \beta_I \cosh\eta_k \sinh\eta_I}{\cosh\eta_k \cosh\eta_I - \beta_I \sinh\eta_k\sinh\eta_I - \beta_I \cos(\phi_k - \phi_I)}\,,
\end{align}
where $(\eta_a, \eta_b)$ specifies the veto region. Integrating first over the angle $\phi_k$ and then over $\eta_k$ yields
\begin{align}\label{eq:JI_result}
J_I=\arcoth\!\left(\frac{\cosh \eta_I \cosh\eta_b- \beta_I \sinh\eta_I \sinh\eta_b}{\sqrt{(\cosh \eta_I \cosh \eta_b- \beta_I \sinh \eta_I \sinh \eta_b)^2-\beta_I ^2}}\right) - (b \to a)\,,
\end{align}
which now depends explicitly on both boundaries, not just the size of the rapidity gap. In the limit of $\beta \xrightarrow{} 1$, we reproduce the massless case~\cite{Becher:2023mtx}
\begin{align}
    J_i=\left(\eta_a- \eta_b\right) \text{sign}\left(\eta_i- \eta_b\right).
\end{align}
The angular integrals $\tilde{J}_\alpha^{IJ}$ appearing in $\boldsymbol{X}^{\Coul^2}$ are more difficult to evaluate. We can, however, relate them to $J_I$ in the case of $2\to t\Bar{t}$ processes, which we will discuss below.

\section{Top-quark pair production cross sections}
\label{sec:top_quark_pair_production}

In this section, we present numerical results for the effect of the new super-leading terms on the partonic gap-between-jet cross section for $t\Bar{t}$ production \cite{ATLAS:2012al,ATLAS:2015yqo,CMS:2017oyi,CMS:2021maw}. From now on, we explicitly consider the ${2\to 2}$ processes $q\Bar{q}\to t\Bar{t}$ and $gg \to t \Bar{t}$. We work in the partonic center-of-mass frame, where these are back-to-back collisions, which greatly simplifies the kinematics.

\subsection{Color space}
\label{subsec:Color_space}

In order to perform the traces $\langle \boldsymbol{\mathcal{H}}_{2 \to 2}(\mu_h) \boldsymbol{X}^{\Coul} \otimes_{\!\int}\!\boldsymbol{1}\rangle$ and $\langle \boldsymbol{\mathcal{H}}_{2 \to 2}(\mu_h) \boldsymbol{X}^{\Coul^2} \otimes_{\!\int}\!\boldsymbol{1}\rangle$, we write the hard functions and the Coulomb operators as matrices in color space. For $q\Bar{q}\to t \Bar{t}$, we use the basis 
\begin{align}\label{eq:color_basis_quarks}
    |\mathcal{D}_1^q\rangle = \delta_{a_2 a_1}\delta_{a_3 a_4}\,, \qquad
    |\mathcal{D}_2^q\rangle = t_{a_2 a_1}^a t_{a_3 a_4}^a\,,
\end{align}
while for $gg \to t\Bar{t}$, we take
\begin{align}\label{eq:color_basis_gluons}
    |\mathcal{D}_1^g\rangle = \delta_{a_1 a_2}\delta_{a_3 a_4}\,, \qquad 
    |\mathcal{D}_2^g\rangle = i f^{a_1 a_2 c}\, t_{a_3 a_4}^c\,, \qquad
    |\mathcal{D}_3^g\rangle = d^{a_1 a_2 c}\, t_{a_3 a_4}^c\,.
\end{align}
Here, $t_{ab}^c$ denote the generators of the fundamental representation and $d^{abc}$ are the totally symmetric structure constants.

\subsubsection{Hard functions}

The spin-averaged hard functions in this basis can be taken from \cite{Ahrens:2010zv} and read (adjusting the prefactors to our conventions) 
\begin{align}\label{eq:hard_function_qqbar}
    \frac{1}{4} \sum_{\text{spins}} \widetilde{\boldsymbol{\mathcal{H}}}_{q\Bar{q} \to t \Bar{t}}(\mu_h) = 16\pi^2 \als^2(\mu_h) \left[\frac{t^2 + u^2}{M^4} + \frac{2 m_t^2}{M^2}\right] \begin{pmatrix}
        0 & 0 \\
        0 & 2 
    \end{pmatrix}
\end{align}
and
\begin{align}\label{eq:hard_function_gg}
    \frac{1}{4} \sum_{\text{spins}} \widetilde{\boldsymbol{\mathcal{H}}}_{gg \to t \Bar{t}}(\mu_h) = 16\pi^2 \als^2(\mu_h) \, \frac{M^4}{2 t u}\left[\frac{t^2 + u^2}{M^4} + \frac{4 m_t^2}{M^2} - \frac{4 m_t^4}{t u}\right] \begin{pmatrix}
        \frac{1}{N_c^2} & \frac{1}{N_c} \frac{t - u}{M^2} & \frac{1}{N_c}\\
        \frac{1}{N_c} \frac{t - u}{M^2} &  \frac{(t - u)^2}{M^4} & \frac{t - u}{M^2} \\
        \frac{1}{N_c} & \frac{t - u}{M^2} & 1
    \end{pmatrix}.
\end{align}
The momentum variables appearing in these expressions are defined as $M^2 = (p_3+p_4)^2$, $t = (p_1 - p_3)^2-m_t^2$ and $u = (p_2 - p_3)^2 - m_t^2$. 

\subsubsection{Color operators}

For the $2\to t\Bar{t}$ processes, the operator $\boldsymbol{X}^\Coul$ \eqref{eq:XCoul} takes the form
\begin{align}\label{eq:XCoul_2_to_ttbar}
    \boldsymbol{X}^\Coul_{2\to t\Bar{t}} = J_{43} v_{34}f^{abe}f^{cde}\tope_1^a \tope_2^b \tope_3^c \tope_4^d\,,
\end{align}
where we have defined $J_{43}=J_4-J_3$.
We obtain the matrix representation for the two processes by using ColorMath \cite{Sjodahl:2012nk} as
\begin{align}
    \boldsymbol{X}^\Coul_{q\Bar{q} \to t \Bar{t}} &= J_{43} v_{34} \frac{1}{8} N_c^3 C_F 
    \begin{pmatrix}
        0 & 1 \\
        1 & 0
    \end{pmatrix},&
    \boldsymbol{X}^\Coul_{gg \to t \Bar{t}} &= J_{43} v_{34} \frac{1}{4} N_c^4 C_F 
    \begin{pmatrix}
        0 & 1 & 0 \\
        1 & 0 & 0 \\
        0 & 0 & 0
    \end{pmatrix}.
\end{align}
For $\boldsymbol{X}^{\Coul^2}$, only the first line in \eqref{eq:C2Coul_trace_evaluated} contributes for the $2\to t\Bar{t}$ processes. It takes the explicit form
\begin{align}\label{eq:X2Coul_2_to_ttbar}
    \boldsymbol{X}^{\Coul^2}_{2\to t\Bar{t}} = v_{34}^2 f^{abe}f^{cde} \left(\tope_3^c \{\tope_4^b,\tope_4^d\}- \tope_4^c \{\tope_3^b,\tope_3^d\}\right)\left(\tilde{J}_1^{34} \tope_1^a + \tilde{J}_2^{34} \tope_2^a \right),
\end{align}
where we have used that $\tilde{J}_\beta^{JI} = - \tilde{J}_\beta^{IJ}$. Using the simplification $\tilde{J}_1^{34}-\tilde{J}_2^{34} = -J_{43}$, we obtain the matrix representations
\begin{align}
    \boldsymbol{X}^{\Coul^2}_{q\Bar{q} \to t \Bar{t}} &=  J_{43} v_{34}^2 \frac{1}{4} N_c^3 C_F 
    \begin{pmatrix}
        0 & 1 \\
        1 & 0
    \end{pmatrix},&
    \boldsymbol{X}^{\Coul^2}_{gg \to t \Bar{t}} &= J_{43} v_{34}^2 \frac{1}{2} N_c^4 C_F 
    \begin{pmatrix}
        0 & 1 & 0 \\
        1 & 0 & 0 \\
        0 & 0 & 0
    \end{pmatrix}.
\end{align}
This shows that $\boldsymbol{X}^{\Coul^2}_{2\to t\Bar{t}} = 2v_{34} \boldsymbol{X}^\Coul_{2\to t\Bar{t}}\,$, which can be also shown at the operator level for the center-of-mass frame.

\subsection{Partonic cross section}
\label{subsec:Partonic_cross_section}

In the center-of-mass frame  of the $t\bar{t}$ pair, the relevant quantities $\beta_I = |\boldsymbol{p}_I| / E_I$ and pseudorapidity $\eta_I = \artanh\!\left(\cos \theta_I\right)$ simplify to $\beta_3 = \beta_4 \equiv \beta$ and $\eta_3 =-\eta_4 \equiv \eta$ for the two heavy quarks.
We can also re-express the kinematical variables appearing in the hard functions in terms of $\beta$ and $\eta$ by observing that
\begin{align}\label{eq:def_kinematical_variables}
    M^2 = \frac{4 m_t^2}{1-\beta^2}\,, \qquad
    t = - \frac{M^2}{2} \left(1- \beta\tanh \eta\right), \qquad
    u = - \frac{M^2}{2} \left(1+ \beta\tanh\eta\right).
\end{align}
Additionally, we can rewrite the kinematical factor $v_{34}$ as
\begin{align}\label{eq:v3v4factor_cms}
    v_{34}=\frac{v_3 \cdot v_4}{\sqrt{(v_3 \cdot v_4)^2-1}} -1 &= \frac{(1-\beta)^2}{2\beta}\,.
\end{align}
Finally, for a $2\to2$ process, the differential cross section of the resummed Coulomb SLLs yields
\begin{align}\label{eq:sigma_SLL_Coul_final_result}
    \left(\frac{d\sigma}{d\eta}\right)^{\text{SLL},\Coul} &= -\frac{1}{\cosh^2 \eta} \frac{\beta}{32\pi M^2} \frac{1}{\mathcal{N}_1 \mathcal{N}_2} \nonumber \\
    & \times \bigg\{16\pi^2\Tr\left(\widetilde{\boldsymbol{\mathcal{H}}}_{2 \to 2}(\mu_h) \boldsymbol{X}^\Coul\right) \, \int_{1}^{x_s}\frac{dx}{x} \frac{1}{\beta_0^3}   U^c(1; \mu_h, \mu) \left(\ln^2 x_s-\ln^2 x \right) \nonumber \\
    &\qquad  +\frac{3}{2}\pi^2 \Tr\left(\widetilde{\boldsymbol{\mathcal{H}}}_{2\to 2}(\mu_h)  \boldsymbol{X}^{\Coul^2}\right)\, \frac{1}{\beta_0^3} \ln^3 x_s\bigg\}\,,
\end{align}
where the multiplicity factors $\mathcal{N}_1$ and $\mathcal{N}_2$ depend on the representation of the partons and $\Tr$ now denotes the usual trace of a matrix. It is important to note that the quark-initiated process does not actually contribute to the cross section. The reason for this is that the color traces vanish for $q \Bar{q} \to t \Bar{t}$ because $\boldsymbol{X}^\Coul$ is antisymmetric under the exchange of the colors of $q$ and $\Bar{q}$ (see relation \eqref{eq:XCoul_2_to_ttbar}), which renders the matrix representations of $\boldsymbol{X}^{\Coul}$ off-diagonal. On the other hand, the hard function for the quark-initiated case turns out to be symmetric under this exchange, yielding a diagonal matrix. The trace of the product of the two vanishes because of these symmetry properties. Therefore, Coulomb SLLs arise exclusively in the gluon-initiated process $gg\to t\Bar{t}$. 
The multiplicity factors then read $\mathcal{N}_1 = \mathcal{N}_2 = 2(N_c^2-1)$.

We stress again that while in the above expressions, the scale integrals are written without accounting for the top-mass threshold, we do include it for the numerical evaluation of the expressions.

\subsection{Sommerfeld resummation of Coulomb phases}
\label{subsec:Sommerfeld_resummation}

The Coulomb-induced SLLs exhibit an interesting feature in the low-$\beta$ region, where the double Coulomb insertion no longer vanishes in the limit $\beta\to 0$ (see the discussion of Figure~\ref{fig: Coulomb betaPlot Single Double Multi eta2} below), even though the phase space closes in this limit. This behavior follows from the fact that each Coulomb insertion scales as $1/\beta$ in the $\beta \to 0$ limit, as encoded in the kinematical factor $v_{34}$. 

This effect is a manifestation of the Sommerfeld enhancement close to threshold \cite{Sommerfeld:1931qaf}. A consistent study of $t\bar{t}$-production requires an all-orders resummation of these divergent terms, which amounts to the resummation of an arbitrary even number of Coulomb phases, corresponding to color traces of the form
$\langle \boldsymbol{\mathcal{H}} (\boldsymbol{V}^\Coul)^{2n} \overline{\boldsymbol{\Gamma}} \otimes_{\!\int}\!\boldsymbol{1} \rangle$. We perform this calculation analogously to Section~\ref{sec:new-sources-of-SLLs} and relegate its details to Appendix~\ref{appendix:resummation_multi_Coulomb}. We find that, when the Coulomb insertions are resummed, the result is well-behaved in the $\beta\to 0$ limit. Note that this requires only the resummation of power-divergences $1/\beta$ corresponding to the Sommerfeld effect. Beyond this, also soft gluon logarithms $\ln\beta$ can appear close to threshold, see e.g.\  \cite{Beneke:1999qg,Beneke:2009rj,Czakon:2009zw,Ahrens:2010zv,
Beneke:2011mq,Ahrens:2011mw,Cacciari:2011hy,Ahrens:2011px}, but these are unrelated to the SLLs considered in this work.

Figure~\ref{fig: Coulomb betaPlot Single Double Multi eta2} shows the Coulomb-SLL contribution to the differential cross section of $gg\to t\bar{t}$, normalized to the Born-level result. We fix the veto region to $\eta \in [-1,1]$ and choose $\eta=2$ for the final-state top, while varying $\beta$. The numerical results depend on the hard and soft scales, where the hard scale is related to $\beta$ through 
\begin{align}\label{eq:betarela}
    \mu_h = \sqrt{\hat s} = \frac{2 m_t}{\sqrt{1-\beta^2}} \,,
\end{align}
which implies that for fixed $\hat s$ the variable $\beta$ is always less than~1. Here and in the following, we set the soft jet-veto scale to $Q_0 =\SI{20}{GeV}$ unless stated otherwise, and use the top mass $m_t = \SI{173.21}{GeV}$. The blue line shows the sum of single- and double-Coulomb insertions, whereas the gray line shows the single-Coulomb insertion for reference. One clearly sees how the double-Coulomb insertion renders the differential cross-section non-vanishing in the low-$\beta$ regime. The red curve then shows the result when Coulomb insertions are resummed to all orders, which exhibits the expected behavior instead. As $\beta$ increases, the red and blue curves approach each other, such that the double-Coulomb insertion approximates the resummed result well for $\beta \gtrsim 0.3$.

\begin{figure}
    \centering
    \includegraphics[]{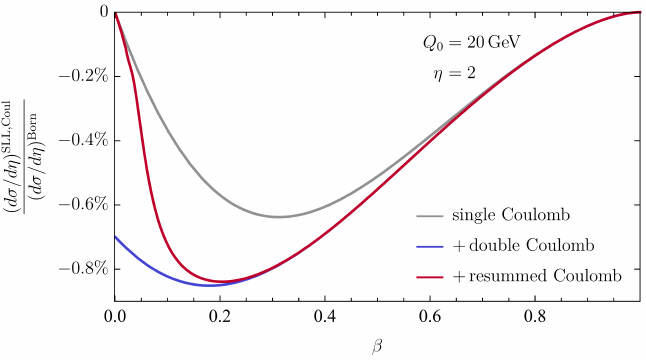}
    \caption{Dependence on $\beta$ of the Coulomb-SLL contributions to the differential cross section of $gg\to t\bar{t}$, normalized to the Born cross section. The jet-veto scale is set to $Q_0 =\SI{20}{GeV}$,    with the pseudorapidity of the top quark chosen as $\eta=2$.
    The gray line indicates the single Coulomb insertion, while the blue line additionally includes the double Coulomb insertion. The red line shows the effect of resumming the Coulomb insertions.}
    \label{fig: Coulomb betaPlot Single Double Multi eta2}
\end{figure}

\subsection{Numerical results}
\label{subsec:Numerical_results}

\begin{figure}[t]
    \centering
    \includegraphics[]{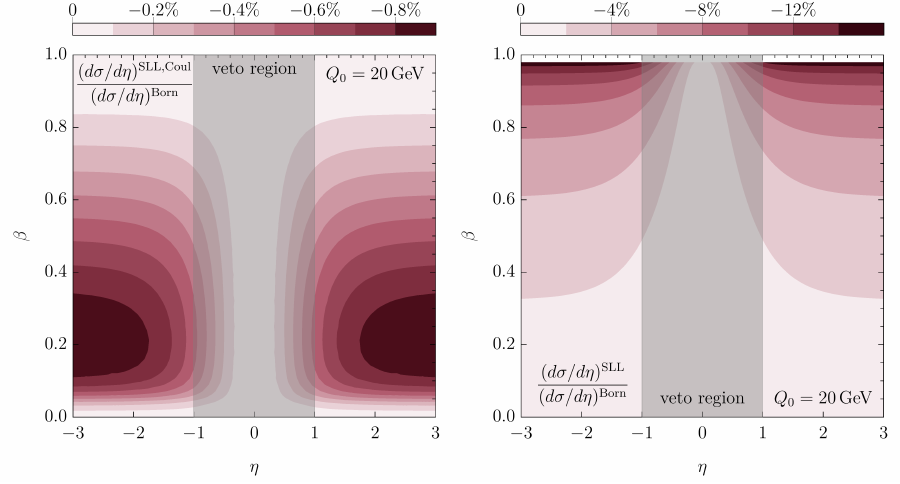}
      \caption{SLL contribution to the differential cross section, normalized to the Born cross section. On the left side we show only the Coulomb-SLL contributions, while in the right the known Glauber SLLs are shown in addition. These plots also include the resummation of an arbitrary even number of Coulomb insertions to correctly treat the Sommerfeld enhancement close to threshold (see Section~\ref{subsec:Sommerfeld_resummation}). The jet-veto scale is fixed to $Q_0 = 20\,\mathrm{GeV}$, while the pseudorapidities of the top quarks are restricted to lie outside the veto region marked in gray.}
    \label{fig: Contour Plot Including Multi Coulomb}
\end{figure}

We are now ready to present the full numerical results for the resummation of SLLs. To this end, we show the regions in $\eta$ and $\beta$ for which the SLLs are most pronounced in Figure~\ref{fig: Contour Plot Including Multi Coulomb}. The left panel illustrates the super-leading logarithmic contributions caused by $\boldsymbol{V}^\Coul$ as discussed above, including the resummation of an arbitrary even number of Coulomb insertions. The hard $t$- and $\bar{t}$-jets are restricted to lie outside the veto region, depicted in gray. (For completeness, we still show the results inside this forbidden region.) Darker colors correspond to contributions of larger negative magnitude. The largest relative contribution is is achieved at $\beta \sim 0.2$.

The complete SLL contributions arising from Glauber and Coulomb insertions are displayed in the right panel of Figure~\ref{fig: Contour Plot Including Multi Coulomb}. While the Glauber contributions are small at low $\beta$, they increase as the hard scale $\mu_h$ grows. For intermediate $\beta$ $(\sim 0.3)$, both contributions are of similar size. Both effects increase moderately as the separation of the top jets in $\eta$ becomes larger.

\begin{figure}
    \centering
    \includegraphics[]{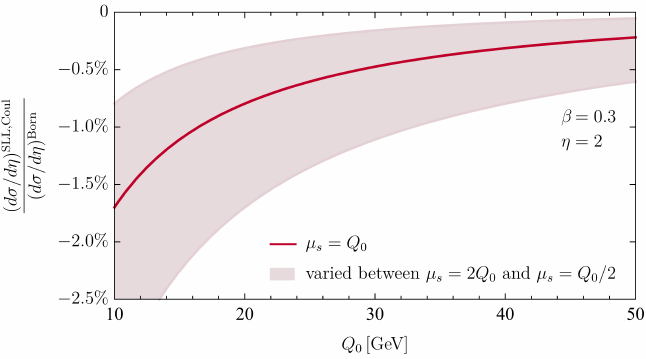}
    \caption{$Q_0$ dependence of the Coulomb SLLs. The shaded band indicates the scale variation $Q_0/2 \leq \mu_s \leq 2Q_0$ as an estimate of the perturbative uncertainties.}
    \label{fig: Coulomb Q0 Including Multiple Coulomb}
\end{figure}

In order to analyze the $Q_0$ dependence of the Coulomb SLLs, we show the effect of varying $Q_0$ in Figure~\ref{fig: Coulomb Q0 Including Multiple Coulomb}. We choose the reference point $\eta = 2$ and $\beta=0.3$ and vary $Q_0$ between 10 and $50\,\text{GeV}$, which results in the red line with the RG-improved treatment of the running coupling. The pink band indicates the variation of the soft scale around $Q_0$ by a factor of 2.

\FloatBarrier

\section{Conclusions}
\label{sec:conclusions}

In this work, we have performed the first resummation of super-leading logarithms (SLLs) for gap-between-jets cross sections in the presence of massive final states. By deriving the soft anomalous dimension including massive quarks, we have identified an additional Coulomb phase as a new source of SLLs: for processes with at least two colored massive particles in the final state, the Coulomb phase generates a tower of SLLs, which we called Coulomb SLLs. The known Glauber SLLs stay unchanged.

As a phenomenological case study, we have analyzed the example of $t\bar t$ production at the LHC. Interestingly, the Coulomb contribution appears only in the gluon-initiated subprocess $gg\to t\bar t$, and not the quark-initiated one. The Sommerfeld enhancement, a divergent behavior of the Coulomb phases in the $\beta\to 0$ limit, necessitates the resummation of these contributions to all orders. 
We observe that for $\beta\sim 0.3$, Coulomb and Glauber contributions are of similar importance, while for large $\beta$ the Glauber contributions dominate.

For the (not-strictly SLL) terms with two or more Coulomb insertions, shown in \eqref{eq:C2Coul_trace_evaluated} and \eqref{eq:X_nCoul}, respectively, no colored initial states are required. This means our framework can be used for a Coulomb resummation in gap-between-jets cross sections at lepton colliders. One could for example capture the effects of radiation cuts in $t\bar{t}$ + jet production.
In contrast to this, the established threshold and Coulomb resummation approaches based on potential non-relativistic QCD are designed for global observables \cite{Beneke:1999qg,Beneke:2010da,Beneke:2013jia,Piclum:2018ndt,Beneke:2024sfa}.

As in previous studies of SLL resummation, the scale uncertainties in our numerical estimates are significant, calling for an improved analysis of subleading corrections. One first step in this direction was the inclusion of the running coupling\cite{Boer:2024hzh}, which we applied to the Coulomb SLLs. For massless final states, another step was achieved by the resummation of arbitrarily many Glauber phases \cite{Boer:2023jsy,Boer:2023ljq,Boer:2024xzy}. In the case of massive final states, one could additionally pursue a resummation of Coulomb insertions in full generality.
Furthermore, the resummation of SLLs in the presence of multiple purely soft emissions remains a challenge \cite{Forshaw:2025fif}. At a later stage, all these advancements should be combined in a phenomenological analysis of hadronic cross sections, extending the work in \cite{Becher:2024nqc}.

\subsubsection*{Acknowledgments}
We thank Michel Stillger and Einan Gardi for useful discussions. This research received funding from the European Research Council (ERC) under the European Union’s Horizon 2022 Research and Innovation Program (ERC Advanced Grant agreement No.101097780, EFT4jets). Views and opinions expressed are those of the authors and do not reflect those of the European Union or the ERC Executive Agency. Neither the European Union nor the granting authority can be held responsible for them. U.B.\ gratefully acknowledges support from the Alexander von Humboldt Foundation and thanks Fermilab for hospitality, where part of this work was carried out. The work reported here was also supported by the Cluster of Excellence \textit{Precision Physics, Fundamental Interactions, and Structure of Matter} (PRISMA$^+$, EXC 2118/1) within the German Excellence Strategy (Project-ID 390831469).

\begin{appendix}
\section{Quasi-collinear enhancement}
\label{appendix:quasi_collinear_enhancement}

Here we discuss the quasi-collinear enhancement that appears in the time-like quasi-collinear splitting $q \to q + g$ of a massive quark (see Section~\ref{subsubsec:collinear-limit}). We parametrize $p_I \cdot p_j = \xi (1-\xi) E_P^2 \,\tilde{n}_I \cdot n_j$ with $\tilde{n}_I\equiv p_I/E_I$ and obtain
\begin{align}\label{eq:collinear_splitting_timelike_qqg_before_evaluation}
    &\langle \boldsymbol{\mathcal{H}}_{m+1}(\argsets{n}{v},\argset{m},\varepsilon) \otimes_{\!\int} \boldsymbol{\mathcal{W}}_{m+1}(\argsets{n}{v},\varepsilon) \rangle \bigg|_{q \to q + g,\; m_q \neq 0} \nonumber \\
    &\qquad  = 4\pi \als \tilde{\mu}^{2 \varepsilon} \langle \prod_{\gamma=3}^{m+1} \int [d\Omega_\gamma] \int\hspace{-9pt}\int\! d\mathcal{E}_m \int_0^1 d\xi \frac{E_P}{\tilde{c}^\varepsilon (2\pi)^2}  \left[\frac{(\xi^2E_P^2-m_q^2)(1-\xi)^2E_P^2}{(E_P^2-m_q^2)}\right]^{(d-3)/2} \nonumber \\
    &\quad\qquad  \times \theta(\xi E_P - m_q)\, \theta((1-\xi) E_P ) \frac{1}{\xi (1-\xi) E_P^2 \,\tilde{n}_I \cdot n_j } \nonumber \\
    &\quad\qquad \times C_F \left[\frac{1+\xi^2}{1-\xi} - \varepsilon(1-\xi) - \frac{m_q^2}{\xi (1-\xi) E_P^2 \,\tilde{n}_I \cdot n_j}\right] {\widetilde{\boldsymbol{\mathcal{H}}}}_{m}(\argset{\hat{p}},\argset{\hat{m}},\mu)  \boldsymbol{\mathcal{W}}_{m}(\argsets{\hat{n}}{\hat{v}},\mu)  \rangle\,.
\end{align}
Next we perform the phase-space integrations related to the additional emitted gluon $j$. The two relevant angular integrals yield
\begin{align}
    \int [d\Omega_j]\frac{1}{(\tilde{n}_I \cdot n_j)} &= \left(1-\frac{m_q^2}{E_I^2}\right)^{-\frac{1}{2}}\artanh\!\left(\sqrt{1-\frac{m_q^2}{E_I^2}}\right) + \mathcal{O}(\varepsilon)\,,  \nonumber \\
    \int [d\Omega_j]\frac{1}{(\tilde{n}_I \cdot n_j)^2} &= \frac{E_I^2}{m_q^2} + \mathcal{O}(\varepsilon)\,. \label{eq:angular_integral_ntilde_n}
\end{align}
The $\theta$-functions restrict the integration domain to $\xi\in\left[m_q/E_P,1\right]$,
such that we still expect a divergence from the region $\xi \to 1$, while the region $\xi\to 0$ is regulated. We can extract the pole of the $\xi$-integration by replacing
\begin{align}\label{eq:collinear_splitting_timelike_extracting_divergence_xi}
    (1-\xi)^{-1+n\varepsilon} = \delta(1-\xi) \frac{1}{n\varepsilon} + \mathcal{O}(\varepsilon^0)\,.
\end{align}
This leads to
\begin{align}\label{eq:divergence_collinear_splitting_timelike_qqg}
    &\langle \boldsymbol{\mathcal{H}}_{m+1}(\argsets{n}{v},\argset{m},\varepsilon) \otimes_{\!\int} \boldsymbol{\mathcal{W}}_{m+1}(\argsets{n}{v},\varepsilon) \rangle \bigg|_{q \to q + g, \; m_q \neq 0} \nonumber \\
    &\quad\!  = \frac{\als}{4\pi} \langle \boldsymbol{\mathcal{H}}_{m}(\argsets{\hat{n}}{\hat{v}},\argset{\hat{m}},\mu) 
    \left\{ - 4 C_F \frac{1}{\varepsilon} \left(1-\frac{m_q^2}{E_P^2}\right)^{\!-\frac{1}{2}}\!\artanh\!\left(\sqrt{1-\frac{m_q^2}{E_P^2}}\right)\! + 2 C_F \frac{1}{\varepsilon} + \mathcal{O}(\varepsilon^0)\! \right\} \nonumber \\
    &\quad\qquad \qquad 
    \otimes_{\!\int}  \boldsymbol{\mathcal{W}}_{m}(\argsets{\hat{n}}{\hat{v}},\mu)  \rangle\,.
\end{align}
However, these divergences are purely soft. They arise from $\xi \to 1$, which corresponds to a soft gluon, while the angular integrals are finite. Since the gluon can never become truly collinear to the massive quark, the $\artanh$-term poses an enhancement but does not lead to a collinear divergence as in the massless case. This contribution is already contained in our discussion of the soft limit. The same applies for the splitting of $\bar{q} \to \bar{q} + g$.
 
For small $m_q/E_P$, expanding the first term in the curly brackets above yields
\begin{align}\label{eq:collinear_splitting_timelike_qqg_reexpand_enhancement}
    & - 4 C_F \frac{1}{\varepsilon} \left(1-\frac{m_q^2}{E_P^2}\right)^{-\frac{1}{2}}\artanh\!\left(\sqrt{1-\frac{m_q^2}{E_P^2}}\right) 
    = 4 C_F \frac{1}{\varepsilon} \ln\!\left(\frac{m_q}{2 E_P}\right) + \mathcal{O}\!\left(\frac{m_q}{E_P}\right) \nonumber\\
    &= 4 C_F \left[ \left( \frac{1}{2\epsilon^2} 
     + \frac{1}{\varepsilon} \ln\!\left(\frac{\mu}{2 E_P}\right) \right) 
     - \left( \frac{1}{2\epsilon^2} 
     + \frac{1}{\varepsilon} \ln\!\left(\frac{\mu}{m_q}\right) \right) \right] + \mathcal{O}\!\left(\frac{m_q}{E_P}\right).
\end{align}
The first term inside the brackets in the last line reproduces the logarithm of the massless case, while the last term is absent for $m_q=0$.

\section{Resummation of multiple Coulomb insertions}
\label{appendix:resummation_multi_Coulomb}

In this appendix, we discuss the resummation of multiple Coulomb insertions, which is relevant in order to obtain the correct behavior for $\beta \to 0$, as discussed in Section~\ref{subsec:Sommerfeld_resummation}. In this analysis, we again focus on $2 \to t \bar{t}$ processes, which simplifies the discussion. 

\subsection{Color structures}

Similar to the case of two Coulomb insertions, where only the first line in \eqref{eq:C2Coul_trace_evaluated} contributes to $2\to t\Bar{t}$ processes, we do not need to consider terms involving more than two heavy particles. We start from the first line in $\boldsymbol{X}^{\Coul^2}$ and subsequently add more Coulomb insertions. We do not give the color-space operators in the following but instead focus on their matrix representations. As for $\boldsymbol{X}^{\Coul^2}$, we factor out $(-4\pi^2)$ in the definition of $\boldsymbol{X}^{\Coul^n}$, i.e.\
\begin{align}\label{eq:X_nCoul}
    \langle \boldsymbol{\mathcal{H}} (\boldsymbol{V}^\Coul)^n \overline{\boldsymbol{\Gamma}} \otimes_{\!\int}\!\boldsymbol{1} \rangle \equiv -4\pi^2 \langle \boldsymbol{\mathcal{H}} \boldsymbol{X}^{\Coul^n} \otimes_{\!\int}\!\boldsymbol{1} \rangle \qquad \text{for} \qquad n\geq 2\,.
\end{align}
Additional factors of $\pi$ arising for $n>2$ are included inside $\boldsymbol{X}^{\Coul^n}$ to simplify the notation. For three Coulomb insertions, we obtain
\begin{align}
    \boldsymbol{X}^{\Coul^3}_{q\Bar{q} \to t \Bar{t}} &= -\frac{1}{8} \pi J_{43} v_{34}^3  N_c^4 C_F 
    \begin{pmatrix}
        0 & i \\
        -i & 0
    \end{pmatrix} , \nonumber \\
    \boldsymbol{X}^{\Coul^3}_{gg \to t \Bar{t}} &= - \frac{1}{4} \pi  J_{43} v_{34}^3  N_c^5 C_F 
    \begin{pmatrix}
        0 & i & 0 \\
        -i & 0 & 0 \\
        0 & 0 & 0
    \end{pmatrix} ,
\end{align}
where $v_{34}=(1-\beta)^2/(2\beta)$ as given in \eqref{eq:v3v4factor_cms}. As expected, the triple Coulomb insertion is purely imaginary and therefore not relevant for QCD processes, which always correspond to real hard functions. In the following, we will therefore not consider the insertion of an odd number of Coulomb phases. However, starting from the expression of $ \boldsymbol{X}^{\Coul^3}$ in terms of color space operators, we can once again add another Coulomb insertion to obtain
\begin{align}
    \boldsymbol{X}^{\Coul^4}_{q\Bar{q} \to t \Bar{t}} &= -\frac{1}{16} \pi^2 J_{43} v_{34}^4  N_c^5 C_F 
    \begin{pmatrix}
        0 & 1 \\
        1 & 0
    \end{pmatrix} 
    = -\frac{1}{4} \pi^2 v_{34}^2 N_c^2 \boldsymbol{X}^{\Coul^2}_{q\Bar{q} \to t \Bar{t}}\, , \nonumber \\
    \boldsymbol{X}^{\Coul^4}_{gg \to t \Bar{t}} &= - \frac{1}{8} \pi^2 J_{43} v_{34}^4  N_c^6 C_F 
    \begin{pmatrix}
        0 & 1 & 0 \\
        1 & 0 & 0 \\
        0 & 0 & 0
    \end{pmatrix}
    = -\frac{1}{4} \pi^2 v_{34}^2 N_c^2 \boldsymbol{X}^{\Coul^2}_{gg \to t \Bar{t}} \,.
\end{align}
This is an important result, since it shows that
\begin{align}
    \boldsymbol{X}^{\Coul^{2n}}_{2\to t\Bar{t}} = \left( v_{34}\frac{i \pi N_c}{2}\right)^{2(n-1)}\! \boldsymbol{X}^{\Coul^2}_{2\to t\Bar{t}} \qquad \forall n \in \mathds{N} \,.
\end{align}
An arbitrary even number of Coulomb insertions can thus be reexpressed as $\boldsymbol{X}^{\Coul^2}$ with a kinematics-dependent prefactor. This allows us to resum these contributions. For the following, we define 
\begin{align}
    v_\text{Coul}\equiv v_{34} \frac{\pi N_c}{2} 
    = \frac{(1-\beta)^2}{2\beta} \frac{\pi N_c}{2} \,.
\end{align}

\subsection{Scale integrals}

Like for the RG-improved resummation approach (see Section~\ref{subsec:scale-int}), we observe an exponentiation of Coulomb phases. However, only even numbers of Coulomb insertions contribute to the cross section. The associated evolution operator takes the form
\begin{align}
   &\boldsymbol{U}_{\text{SLL}}^{\text{multi-\Coul}}(\mu_h, \mu_s) = 
    \sum_{n=1}^\infty 
    \int_{\mu_s}^{\mu_h} \frac{d\mu_1}{\mu_1} \gcuspals{\mu_1} \boldsymbol{V}^\Coul
    \int_{\mu_s}^{\mu_1} \frac{d\mu_2}{\mu_2} \gcuspals{\mu_2} \boldsymbol{V}^\Coul 
    \ldots \nonumber \\
   &\qquad \times 
    \int_{\mu_s}^{\mu_{2n-1}} \frac{d\mu_{2n}}{\mu_{2n}} \gcuspals{\mu_{2n}} \boldsymbol{V}^\Coul 
    \int_{\mu_s}^{\mu_{2n}} \frac{d\mu_{2n+1}}{\mu_{2n+1}} \frac{\als(\mu_{2n+1})}{4\pi}\, \overline{\boldsymbol{\Gamma}} \nonumber \\
   &= \sum_{n=1}^\infty \frac{1}{(2n)!} 
    \int_{\mu_s}^{\mu_h} \frac{d\mu}{\mu} \frac{\als(\mu)}{4\pi} 
    \left[ \int_\mu^{\mu_h} \frac{d\mu'}{\mu'} \gcuspals{\mu'} \right]^{2n} 
    \left(\boldsymbol{V}^\Coul\right)^{2(n-1)}
    \left(\boldsymbol{V}^\Coul\right)^2 \overline{\boldsymbol{\Gamma}} ,
\end{align}
where in the last step we have inverted the order of the nested integrations. Since $\left(\boldsymbol{V}^\Coul\right)^2 \overline{\boldsymbol{\Gamma}}$ is an eigenvector to a subsequent application of  $(\boldsymbol{V}^\Coul)^2$ in the trace with the hard function, the operator $\boldsymbol{V}^\Coul$ can be replaced by $v_\text{Coul}$. Performing the remaining integrations at lowest order in RG-improved perturbation theory, we find (for fixed $n_f$) 
\begin{align}\label{eq:Sommerfeld resummed}
    &\langle \boldsymbol{\mathcal{H}}_{2\to M}(\mu_h) \boldsymbol{U}_{\text{SLL}}^{\text{multi-\Coul}}(\mu_h,\mu_s) \otimes_{\!\int}\!\boldsymbol{1} \rangle  \nonumber\\
    &= - 4\pi^2 \langle \boldsymbol{\mathcal{H}}_{2\to M}(\mu_h)  \boldsymbol{X}^{\Coul^2} \otimes_{\!\int}\!\boldsymbol{1} \rangle \left[
    \frac{1}{v^2_\text{Coul}}\frac{1}{2\beta_0}\ln x_s -
    \frac{1}{\gamma_0 v^3_\text{Coul}}
    \sin\left(\frac{\gamma_0}{2\beta_0}v_\text{Coul}\ln x_s \right)
    \right] ,
\end{align}
with $x_s=\alpha_s(\mu_s)/\alpha_s(\mu_h)$. In the fixed-coupling approximation, $\frac{1}{2\beta_0}\ln x_s$ is replaced by $\frac{\als(\Bar{\mu})}{4\pi}\ln\big(\frac{\mu_h}{\mu_s}\big)$. Expanding this expression in a power series in $v_\text{Coul}$, one finds that the leading term agrees with the result shown in \eqref{eq:trace_H_USLL2Coul}. In the limit $\beta\to 0$, the quantity $v_\text{Coul}$ diverges with $1/\beta$, and so do all terms in the series expansion. Yet, the resummed expression in \eqref{eq:Sommerfeld resummed} vanishes $\sim\beta^2$. 

Note that while we only discussed even numbers of Coulomb insertions, since $\boldsymbol{X}^{\Coul^3}$ is known, the same procedure can easily be implemented for an arbitrary odd number of insertions. 

\end{appendix}

\clearpage
\pdfbookmark[1]{References}{Refs}
\bibliography{refs.bib}

\end{document}